\documentclass[twocolumn,floatfix,aps,prd,showpacs]{revtex4}

\usepackage{graphicx}
\usepackage{dcolumn}
\usepackage{bm}
\usepackage{epsf}

\draft

\newcommand{\beq}{\begin{equation}}
\newcommand{\eeq}{\end{equation}}
\newcommand{\beqn}{\begin{eqnarray}}
\newcommand{\eeqn}{\end{eqnarray}}

\newcommand{\pa}{\partial}

\newcommand{\varep}{\varepsilon}

\def\agt{\mathrel{\raise.3ex\hbox{$>$}\mkern-14mu\lower0.6ex\hbox{$\sim$}}}
\def\alt{\mathrel{\raise.3ex\hbox{$<$}\mkern-14mu\lower0.6ex\hbox{$\sim$}}}

\begin{document}

\title{Evolution of neutron stars with toroidal magnetic fields: 
Axisymmetric simulation in full general relativity}

\author{Kenta Kiuchi$^1$}
\author{Masaru Shibata$^2$}
\author{Shijun Yoshida$^3$}

\affiliation{
$^{1}$Department of Physics, Waseda University, 3-4-1 Okubo, 
 Shinjuku-ku, Tokyo 169-8555, Japan \\
$^2$Graduate School of Arts and Sciences, University of Tokyo, 
Komaba, Meguro, Tokyo 153-8902, Japan\\
$^3$Astronomical Institute, Tohoku University, Sendai 980-8578, Japan}

\begin{abstract}
We study the stability of neutron stars with toroidal magnetic fields
by magnetohydrodynamic simulation in full general relativity under
assumption of axial symmetry.  Nonrotating and rigidly rotating
neutron stars are prepared for a variety of magnetic field
configuration.  For modeling the neutron stars, the polytropic
equation of state with the adiabatic index $\Gamma=2$ is used for
simplicity.  It is found that nonrotating neutron stars are
dynamically unstable for the case that toroidal magnetic field
strength varies $\propto \varpi^{2k-1}$ with $k\geq 2$ (here $\varpi$
is the cylindrical radius), whereas for $k=1$ the neutron stars are
stable. After the onset of the instability, unstable modes grow
approximately in the Alfv\'en time scale and, as a result, a
convective motion is excited to change the magnetic field profile
until a new state, which is stable against axisymmetric perturbation,
is reached. We also find that rotation plays a role in stabilization,
although the instability still sets in in the Alfv\'en time scale when
the ratio of magnetic energy to rotational kinetic energy is larger
than a critical value $\sim 0.2$.  Implication for the evolution of
magnetized protoneutron stars is discussed.
\end{abstract}
\pacs{04.25.Dm, 04.40.Nr, 47.75.+f, 95.30.Qd}

\maketitle

\section{Introduction}

Neutron stars observed in nature are magnetized with the typical
magnetic field strength $\sim 10^{11}$--$10^{13}$ G \cite{Lyne}.  The field
strength is often much larger than the canonical value as $\sim
10^{15}$ G for a special class of the neutron stars such as magnetars
\cite{WD}. The field strength at the birth of neutron stars may be
also much larger than the canonical value, because in the
supernova gravitational collapse, rapid and differential rotation of
the collapsing core could amplify the magnetic field. In the presence 
of a radial magnetic field $B^{\varpi}$, the toroidal field $B^T$ 
is amplified by winding in the presence of differential rotation, 
and the field strength increases with time approximately according to 
(see, e.g., \cite{AHM,SLSS})
\beqn
B^T & \sim & B^{\varpi} \Omega t \nonumber \\
&=& 10^{15} 
\biggl({B^{\varpi} \over 10^{12}~{\rm G}}\biggr)
\biggl({\Omega \over 10^2~{\rm rad/s}}\biggr)
\biggl({t \over 10~{\rm s}}\biggr) ~{\rm G},
\eeqn
where we adopt the typical magnitude of the angular velocity and the
typical cooling time of the protoneutron star for $\Omega$ and
$t$. The large field strength of the magnetars may be generated by
such process and subsequently be confined inside the neutron star for
thousands of years \cite{Spruit}. This suggests that even for the
normal pulsar, the toroidal field strength inside the neutron star may
be much larger than the canonical value. Thus, strongly magnetized
neutron stars may be common in nature. In particular, the toroidal
field is likely to be much stronger than the poloidal fields inside 
neutron stars.  In this paper, we focus on the effect of such strong
toroidal magnetic fields for the dynamical evolution of neutron stars.

Stars with purely toroidal magnetic fields in a stably stratified
structure are known to be unstable against the Tayler instability
\cite{Tayler,Acheson,Tayler2,Spruit99} (see also Appendix A).
According to a perturbative study in
\cite{Tayler,Acheson,Tayler2,Spruit99}, the most unstable motions are
driven by axisymmetric ($m=0$) and nonaxisymmetric $m=1$ modes with
nearly horizontal displacement. The unstable modes are predicted to
grow approximately on an Alfv\'en time scale. The Alfv\'en time scale
of magnetized neutron stars are estimated to be very short as 
\beqn
\tau_A &\sim& {R \over v_A} \nonumber \\
&\sim&30 \biggl({R \over 10~{\rm km}}\biggr)
\biggl({\rho \over 10^{14}~{\rm g/cm^3}}\biggr)^{{1\over 2}}
\biggl({B^T \over 10^{15}~{\rm G}}\biggr)^{-1}~{\rm ms}, \label{taualf}
\eeqn
where $R$ and $\rho$ are characteristic radius and density of the
neutron star, $v_A$ is the Alfv\'en speed, and we use
$v_A=B^T/(4\pi\rho)^{1/2}$.  The time scale for the growth of 
the Tayler instability
is only by one order of magnitude longer than the dynamical time scale
of neutron stars which is $\sim 1$ ms. Thus, the instability associated with
the strong magnetic fields may affect even early evolution of the
protoneutron star and, consequently, supernova explosion.  However,
perturbative studies do not clarify anything in the nonlinear
evolution stage reached after a sufficient growth of the instability.

To understand roles of strong toroidal fields on the evolution
of neutron stars and protoneutron stars, numerical simulation is
probably the best approach. In this paper, we present our new
numerical results obtained by general relativistic magnetohydrodynamic
(GRMHD) simulation, for which our GRMHD code recently
developed \cite{SS05} is used. We prepare neutron stars with purely toroidal
magnetic fields in axisymmetric equilibria computed by a method 
described in \cite{KY}. As a
first step toward a deep understanding of the Tayler instability, we
focus on the $m=0$ mode imposing axial symmetry. As shown
in \cite{Spruit99}, neutron stars are unstable if certain condition is
satisfied for the magnetic field profile and for the rotation rate. We
confirm this fact in the present numerical simulation.  In addition,
we follow evolution of the unstable stars after the onset of the Tayler
instability and show that associated with the growth of this 
instability, a convective motion is driven inside the neutron
star. Then, the magnetic fields are redistributed and eventually their 
profile relaxes to a new state which is stable against axisymmetric
perturbation.

The remainder of this paper is organized as follows. In Sec.~II, we
briefly review formulation and numerical methods for our GRMHD
simulations. Section III presents numerical results for nonrotating
and rotating neutron stars separately.  Section IV is devoted to a
summary and discussion about implication of the present results on the
evolution of neutron stars. In Appendix A, we present a result of
linear perturbative study for neutron stars of purely toroidal
magnetic fields, which validates our numerical results qualitatively.
Throughout this paper, we adopt geometrical units in which $G=1=c$,
where $G$ and $c$ denote the gravitational constant and speed of
light, respectively. Cartesian coordinates are denoted by $x^k=(x, y,
z)$. The coordinates are oriented so that the symmetric axis is along
the $z$-direction. We define the coordinate radius
$r=\sqrt{x^2+y^2+z^2}$, cylindrical radius $\varpi=\sqrt{x^2+y^2}$,
and azimuthal angle $\varphi=\tan^{-1}(y/x)$. Coordinate time is
denoted by $t$. Greek indices $\mu, \nu, \cdots$ denote spacetime
components, and small Latin indices $i, j, \cdots$
denote spatial components.

\section{Method for numerical simulation}

\subsection{Formulation and methods}

The stability of magnetized neutron stars and the fate of unstable
neutron stars are investigated by GRMHD simulation assuming that the
ideal MHD condition holds.  In this paper, we assume the axial
symmetry and focus only on the Tayler instability against axisymmetric
perturbation.  The simulation is performed by a GRMHD code for which
the details are described in \cite{SS05}.  This code makes long-term
numerical evolutions of relativistic magnetized neutron stars
possible.  It solves the Einstein-Maxwell-MHD system of coupled
equations, both in axial symmetry and in 3+1 dimensions, without
approximation.  The code evolves the spacetime metric using the BSSN
formulation~\cite{BSSN}; we evolve the conformal three metric $\tilde
\gamma_{ij}=\gamma^{-1/3}\gamma_{ij}$, a conformal factor
$\phi=\ln(\gamma)/12$, a tracefree extrinsic curvature $\tilde
A_{ij}=e^{-4\phi}(K_{ij}-\gamma_{ij}K_k^{~k}/3)$, trace of the
extrinsic curvature $K_k^{~k}$, and an auxiliary three variable
$F_i=\sum_j \pa_j \tilde \gamma_{ij}$. Here, $\gamma_{ij}$ is the
three-metric and $\gamma={\rm det}(\gamma_{ij})$. For axisymmetric 
simulation, the Cartoon method is employed \cite{alcu,S03}: Namely, 
the Einstein equation is solved in the Cartesian coordinates 
imposing an axisymmetric boundary condition and 
the hydrodynamic equation is in the cylindrical coordinates. 

As in previous axisymmetric simulations (e.g., \cite{S2d,SS05}), 
the following dynamical gauge condition is employed 
\beqn
&&\pa_t \alpha=-\alpha K_k^{~k},\\
&&\pa_t \beta^i=\tilde \gamma^{ij} (F_j +\Delta t \pa_t F_j),
\eeqn
where $\alpha$ is the lapse function, $\beta^i$ the shift vector, 
and $\Delta t$ time step in numerical computation. 

A conservative shock-capturing scheme is employed to integrate the
GRMHD equations. Specifically we use a high-resolution central scheme
\cite{KT,lucas} with the third-order piece-wise parabolic
interpolation and with a steep min-mod limiter in which the limiter
parameter $b$ is set to be 2.5 (see appendix A of \cite{S03}).
Multiple tests have been performed with these codes, including MHD
shocks, MHD wave propagation, magnetized Bondi accretion, and
magnetized accretion onto a neutron star \cite{SS05}.  This code has
been already applied to the evolution of magnetized hypermassive
neutron stars to a black hole \cite{DLSSS,DLSSS2} and to supernova
gravitational collapse of strongly magnetized and rotating core
\cite{SLSS}, and derived reliable numerical results.

In the present paper, we initially give a purely toroidal magnetic
field.  In such a case, poloidal magnetic fields are never generated
in the axisymmetric spacetime. Thus, we only solve the toroidal 
field component. 

As initial conditions for the numerical simulation, we prepare
magnetized neutron stars in equilibrium \cite{KY}. For computing the
equilibrium, we give the polytropic equation of state as
\beqn
P=\kappa \rho^{\Gamma},
\eeqn
where $P$, $\rho$, $\kappa$, and $\Gamma$ are the pressure, 
rest-mass density, polytropic constant, and adiabatic constant. 
In this work, we choose $\Gamma=2$. Because $\kappa$ is arbitrarily 
chosen or else completely scaled out of the problem, we adopt 
the units of $\kappa=1$ in the following (i.e., the units of 
$c=G=\kappa=1$). 

In numerical simulation, we adopt the $\Gamma$-law equation of state 
\beqn
P=(\Gamma-1)\rho \varep,
\eeqn
where $\varep$ is the specific internal thermal energy.

\begin{table*}[t]
\begin{center}
\caption{List of characteristic quantities for neutron stars with
  toroidal magnetic fields. Value of $k$, central density, $\rho_c$,
  baryon rest mass, $M_*$, ADM mass, $M$, ratio of equatorial
  circumferential radius $R$ to $M$, ratio of the rotational kinetic
  energy to the gravitational potential energy, $T_{\rm rot}/W$, ratio of the
  internal thermal energy to $W$, $E_{\rm int}/W$, ratio of
  the electromagnetic energy to $W$, $E_{\rm EM}/W$ ,
  non-dimensional angular momentum parameter, $J/M^2$, central value of
  the lapse function, $\alpha_c$, angular velocity, $\Omega$, and 
  Alfv\'en time scale defined by Eq. (\ref{Alf}). All the quantities
  are shown in units of $c=G=\kappa=1$. For models AXY, BXY, and CXY,
  $k=1$, 2, and 3, respectively. ``X'' denotes the value of $10\rho_c$
  and ``Y''(=H, M, L) denotes the relative strength of the magnetic
  field. Models RAXY and RBXY denote rapidly rotating neutron stars
  (meaning of A, B, X, and Y is the same as above).  Models MBXy
  denote moderately rapidly rotating neutron stars.  Models EBXy and
  SBXy denote rotating models with very strong and strong magnetic
  fields, respectively. ``y'' approximately denotes $100T/W$. In the
  last column, the stability determined by the numerical simulation is
  described.}
\begin{tabular}{cccccccccccccc} \hline
~Model~ & ~$k$~ & $\rho_c$ & $M_*$ & $M$ & $R/M$ 
& ~$T_{\rm rot}/W$~ & ~$E_{\rm int}/W$~ 
& ~$E_{\rm EM}/W$~ & ~~$J/M^2$~~ & $\alpha_c$ & $\Omega$ & $\bar \tau_A/M$ 
& Stable ? \\ \hline
A3H & 1 & ~0.3000~ & ~0.1793~ & ~0.1637~ & ~4.821~ & 0 & 0.5858 & 0.0146 
& 0  & ~0.4786~ & 0 & 67.8 & Yes \\ 
A3L & 1 & 0.3000 & 0.1798 & 0.1637 & 4.749 & 0 & 0.5931 & $9.8\times 10^{-4}$
&0  & 0.4756 & 0 & 256 & Yes \\ 
A2H & 1 & 0.2000 & 0.1715 & 0.1574 & 5.599 & 0 & 0.5148 & 0.0135 
& 0  & 0.5732 & 0 & 89.6 & Yes \\ 
B3H & 2 & 0.3000 & 0.1794 & 0.1637 & 4.783 & 0 & 0.5880 & 0.0122 
& 0  & 0.4772 & 0 & 73.4&No \\ 
B3M & 2 & 0.3000 & 0.1797 & 0.1637 & 4.750 & 0 & 0.5927 & $2.05 \times 10^{-3}$
&0  & 0.4757 & 0 & 177 &No \\ 
B3L & 2 & 0.3000 & 0.1798 & 0.1637 & 4.747 & 0 & 0.5932 & $9.3 \times 10^{-4}$ 
&0& 0.4755 & 0 & 263 &No \\ 
B2H & 2 & 0.2000 & 0.1714 & 0.1573 & 5.543 & 0 & 0.5168 & 0.0108 
& 0  & 0.5715 & 0 & 98.5 &No \\ 
B2M & 2 & 0.2000 & 0.1717 & 0.1573 & 5.516 & 0 & 0.5201 &$3.11 \times 10^{-3}$
&0& 0.5703 & 0 & 182 &No \\ 
B2L & 2 & 0.2000 & 0.1717 & 0.1574 & 5.505 & 0 & 0.5210 & $1.16 \times 10^{-3}$
&0& 0.5700 & 0 & 297 &No \\ 
C3L & 3 & 0.3000 & 0.1798 & 0.1638 & 4.747 & 0 & 0.5931 & $1.16\times 10^{-3}$
&0& 0.4755 & 0 & 236 &No \\ 
C2H & 3 & 0.2000 & 0.1714 & 0.1573 & 5.537 & 0 & 0.5169 & 0.0108 
& 0 & 0.5713 & 0 & 98.5 &No \\ \hline 
RA2H & 1 & 0.2000 & 0.1986 & 0.1821 & 6.732 & 0.0808 & 0.4555 & 0.0145 
& 0.5667  & 0.5383 & ~0.3159~ & 99.5 &Yes \\ 
RA2L & 1 & 0.2000 & 0.2021 & 0.1848 & 6.491&0.0866 &0.4577&$1.78\times 10^{-3}$ 
& 0.5894  & 0.5318 & 0.3271 & 272 &Yes \\ 
RB2S & 2 & 0.2000 & 0.1981 & 0.1817 & 6.283 & 0.0796 & 0.4559 & 0.0176 
& 0.5614  & 0.5382 & 0.3144 & 84.4 &No \\ 
RB2H & 2 & 0.2000 & 0.2002 & 0.1835 & 6.594 & 0.0839 & 0.4549 & 0.0126 
& 0.5783  & 0.5352 & 0.3211 & 104 &Yes \\ 
RB2L & 2 & 0.2000 & 0.2023 & 0.1850 &6.478&0.0869&0.4575 &$1.78\times 10^{-3}$
& 0.5906  & 0.5315 & 0.3276 & 271 & Yes \\ 
MB2H & 2 & 0.2000 & 0.1908 & 0.1751 & 5.863 & 0.0611 & 0.4708 & 0.0163
& 0.4870  & 0.5460 & 0.2852 & 82.6 &No \\ 
MB2M & 2 & 0.2000 & 0.1906 & 0.1747 & 5.797 & 0.0597 & 0.4744 & 0.0108 
& 0.4817  & 0.5456 & 0.2832 & 99.9 &No \\ 
MB2L & 2 & 0.2000 & 0.1903 & 0.1743 & 5.738&0.0581&0.4780&$5.66 \times 10^{-3}$ 
& 0.4758 & 0.5453 & 0.2809 & 137 & Yes \\
MB2L' & 2 & 0.2000 & 0.1856 & 0.1700 &5.649&0.0447&0.4882&$4.66 \times 10^{-3}$
& 0.4151  & 0.5512 & 0.2512 & 149 & Yes \\ 
EB27 & 2 & 0.200  & 0.1915 & 0.1765 & 6.251 & 0.0659 & 0.4550 & 0.0427 
& 0.5040  & 0.5492 & 0.2886 &54.8 &No \\ 
EB25 & 2 & 0.200  & 0.1858 & 0.1713 & 5.995 & 0.0502 & 0.4667 & 0.0433 
& 0.4353  & 0.5559 & 0.2592 &52.6 &No \\ 
EB23 & 2 & 0.2000 & 0.1793 & 0.1653 & 5.798 & 0.0302 & 0.4833 & 0.0392 
& 0.3343  & 0.5635 & 0.2078 & 53.9 &No \\ 
EB21 & 2 & 0.2000 & 0.1745 & 0.1610 & 5.744 & 0.0144 & 0.4922 & 0.0443 
& 0.2286  & 0.5708 & 0.1465 & 50.6 &No\\ 
SB24 & 2 & 0.2000 & 0.1845 & 0.1692 & 5.702 & 0.0430 & 0.4854 & 0.0137 
& 0.4052  & 0.5536 & 0.2459 & 88.0 &No \\ 
SB23 & 2 & 0.2000 & 0.1805 & 0.1656 & 5.629 & 0.0305 & 0.4955 & 0.0111 
& 0.3400  & 0.5586 & 0.2109 & 97.2 &No \\ 
SB21 & 2 & 0.2000 & 0.1744 & 0.1601 & 5.572 & 0.0108 & 0.5091 & 0.0116 
& 0.2005  & 0.5671 & 0.1286 & 95.1 &No \\ 
\hline
\end{tabular}
\end{center}
\end{table*}

\subsection{Diagnostics}

We monitor the total baryon rest mass $M_*$, ADM
(Arnowitt-Deser-Misner) mass $M$, and angular momentum $J$, which are 
computed, in axial symmetry, by
\beqn
&& M_*=\int \rho u^t \sqrt{-g} d^3 x,\\
&& M =\int \rho_{\rm ADM} \sqrt{\gamma} d^3 x,\\
&&  J=\int \rho h u^t u_{\varphi} \sqrt{-g} d^3 x,
\eeqn
where $u^{\mu}$ is the four velocity, $g$ is the determinant 
of the spacetime metric, $h$ is the specific enthalpy 
($h=1+\varep+P/\rho$), and
\beqn
&&\rho_{\rm ADM} = [\rho h (\alpha u^t)^2 - P]e^{-\phi} \nonumber \\
&&~~~~~~~~ + {e^{-\phi} \over 16\pi} \biggl[K_{ij}K^{ij}-(K_k^{~k})^2-
\tilde R e^{-4\phi}\biggr]. 
\eeqn
Here, $\tilde R$ is the Ricci scalar with respect to $\tilde \gamma_{ij}$. 
Hereafter, initial ADM mass is denoted by $M_0$. 

In addition to the above quantities, we 
monitor the internal thermal energy $E_{\rm int}$,
rotational kinetic energy
$T_{\rm rot}$, total kinetic energy $T_{\rm kin}$, and 
electromagnetic energy $E_{\rm EM}$, written by 
\begin{eqnarray}
E_{\rm int} &=& \int \rho u^t \varepsilon \sqrt{-g} d^3x , \\
T_{\rm rot}    &=& {1\over 2}\int \rho h u^t u_{\varphi} \Omega \sqrt{-g} 
d^3x, \label{eq:Trot} \\
T_{\rm kin}    &=& {1\over 2}\int \rho h u^t u_{i} v^i \sqrt{-g} 
d^3x, \label{eq:Tkin} \\
E_{\rm EM}     &=& {1 \over 2} \int b^2 u^t \sqrt{-g} d^3x,
\end{eqnarray}
where $b^2=b^{\mu} b_{\mu}$, $b^{\mu}$ is a magnetic vector 
in the frame comoving with fluid elements (e.g., \cite{SS05}),  
and $v^i=u^i/u^t$. In this paper, magnetic field strength is 
defined by $\sqrt{4\pi b^2}$. We note that $E_{\rm EM}$ is defined 
originally by 
\beqn
E_{\rm EM}=\int T_{\rm EM}^{\mu\nu} n_{\mu} u_{\nu} \sqrt{\gamma} d^3x,
\eeqn
where $T_{\rm EM}^{\mu\nu}$ is the electromagnetic part of 
the energy momentum tensor and $n^{\mu}$ is the hypersurface normal, 
and hence, the definition is different from that in \cite{DLSSS2}. 
This definition is based on mimicking the definition of $E_{\rm int}$, 
which is 
\beqn
E_{\rm int}=\int T_{\rm hydro}^{\mu\nu} n_{\mu} u_{\nu} \sqrt{\gamma} d^3x
-M_*,
\eeqn
where $T_{\rm hydro}^{\mu\nu}(=\rho h u^{\mu} u^{\nu} + Pg^{\mu\nu})$ 
is the non-electromagnetic part of the energy momentum tensor. 

Once each energy component is obtained, 
gravitational potential energy is defined by 
\beqn
W=M_*+E_{\rm int}+E_{\rm EM}+T_{\rm kin}-M. 
\eeqn

The Alfv\'en speed in relativity is defined by
\beqn
v_A=\sqrt{{b^2 \over \rho h + b^2}}.
\eeqn
Associated Alfv\'en time scale is 
\beqn
\tau_A = {L \over v_A},
\eeqn
where $L$ is a characteristic length scale and in the present context,
it is approximately equal to stellar radius.  Usually, the
Alfv\'en time scale denotes a characteristic time during which
Alfv\'en waves propagate for the characteristic length scale. In the
present case, we assume the axial symmetry and presence of purely 
toroidal magnetic field, and hence, Alfv\'en waves play no role. 
Nevertheless, a dynamical instability analyzed in this paper grows 
in the time scale of order $\tau_A$. For this reason, we here define 
an averaged Alfv\'en time scale from global quantities as 
\beqn
&& \bar v_A  \equiv 
\Biggl[{\displaystyle \int b^2 u^t \sqrt{-g} d^3x \over 
\displaystyle \int (\rho h + b^2) u^t \sqrt{-g} d^3x}\Biggr]^{{1 \over 2}}
\nonumber \\
&&~~~~= \sqrt{2 E_{\rm EM} \over M_* + \Gamma E_{\rm int} + 2 E_{\rm EM}}, 
\eeqn
where we use the relation $h=1+\Gamma \varep$ which holds 
in the $\Gamma$-law equation of state. Then, we define 
averaged Alfv\'en time scale as
\beqn
\bar \tau_A = {R \over \bar v_A}, \label{Alf}
\eeqn
where $R$ is equatorial stellar radius. 

\subsection{Initial condition}

We prepare a variety of neutron stars in equilibrium changing the
compactness, profile and strength of toroidal magnetic fields, and
rotational kinetic energy.  The initial conditions are derived in the
same method as that described in \cite{KY}. 
In the present case, we give the toroidal magnetic field according to 
the relation 
\beqn
b_{\varphi}=B_0 u^t (\rho h \alpha^2 \gamma_{\varphi\varphi})^k,
\eeqn
where $k$ and $B_0$ are constants which determine the field profile
and field strength, respectively. Because of the regularity condition 
along the symmetric axis, $k$ has to be a positive integer. 
$\gamma_{\varphi\varphi}$ is the
$\varphi\varphi$ component of $\gamma_{ij}$ and
approximately proportional to $\varpi^2$ near the symmetric axis.
Because $b_{\varphi}$ is a function of $\rho$, the magnetic field is
confined inside the neutron star. 

In this work, we choose $k=1$, 2, and 3. Because $b_{\varphi}$ is
proportional to $\varpi^{2k}$ near the symmetric axis, the toroidal
field strength defined by $B^T \equiv b_{\varphi}
\gamma_{\varphi\varphi}^{-1/2}/\sqrt{4\pi}$ is proportional to
$\varpi^{2k-1}$. Namely, for small values of $k$, the fields are
confined near the symmetric axis.  References
\cite{Tayler,Acheson,Tayler2,Spruit} (and also Appendix A) predict
that stars with $k=1$ are stable against axisymmetric perturbation,
whereas those with $k \geq 2$ are unstable, although rotation could
stabilize the unstable mode. 

Several key quantities which characterize the magnetized neutron stars
are listed in Table I. $B_0$ is chosen so as to get $10^{-3} \alt
E_{\rm EM}/W \alt 4 \times 10^{-2}$.  For typical neutron stars of
mass $\sim 1.4M_{\odot}$, $W \sim 6 \times 10^{53}$ ergs.  The
electromagnetic energy is approximately written as $E_{\rm EM} \sim
(B^T)^2R^3/3$, and hence, the magnetic field strength we consider here
is extremely large as $10^{16}$--$10^{17}$ G for $R \approx 10$ km.
Such choice is done simply to save computational time (note that the
time scale for the growth of unstable modes is proportional to
$(B^T)^{-1}$; see below).  In all the cases, the magnetic field is
strong but not strong enough to modify the stellar structure
significantly; e.g., for the nonrotating case, the shape of the
neutron stars is approximately spherical.

Even from this extreme setting, we can derive a generic physical
essence because scaling relation, associated with the magnetic field
strength, holds for the evolution of the unstable neutron
star. Namely, if the magnetic field strength becomes half, the growth
time scale for the Tayler instability becomes approximately twice
longer, although the qualitative properties about the evolution of the
unstable star are essentially the same. Hence, the artificial choice
of the large magnetic field is acceptable for deriving generic
physical properties.

Compactness of the neutron stars is determined from the conditions that
the central density, $\rho_c$, is 0.300 or 0.200 (in units of
$\kappa=1$).  We note that the maximum rest mass and gravitational
mass of spherical neutron stars for $\Gamma=2$ are 0.1799 and 0.1637,
and the corresponding central density is $\approx 0.318$. This implies
that the nonrotating neutron stars with $\rho_c=0.300$ are close to
the marginally stable point against gravitational collapse. Indeed,
the rest mass and ADM mass for such models are close to the values of
marginally stable stars (cf. Table I). We will show that our code can
follow such extremely compact stars stably for a long time $\agt
3000M_0$. By contrast, with $\rho_c=0.2$, the ratio of the stellar
radius to the ADM mass becomes $\sim 5.5$ for the nonrotating neutron
stars, which is a typical magnitude for neutron stars (for a 
hypothetical value of ADM mass $1.35M_{\odot}$, $R \approx 11$ km).  Thus, in
this paper, we consider very compact and reasonably compact neutron
stars.

\begin{figure*}[th]
\epsfxsize=3.in
\leavevmode
\epsffile{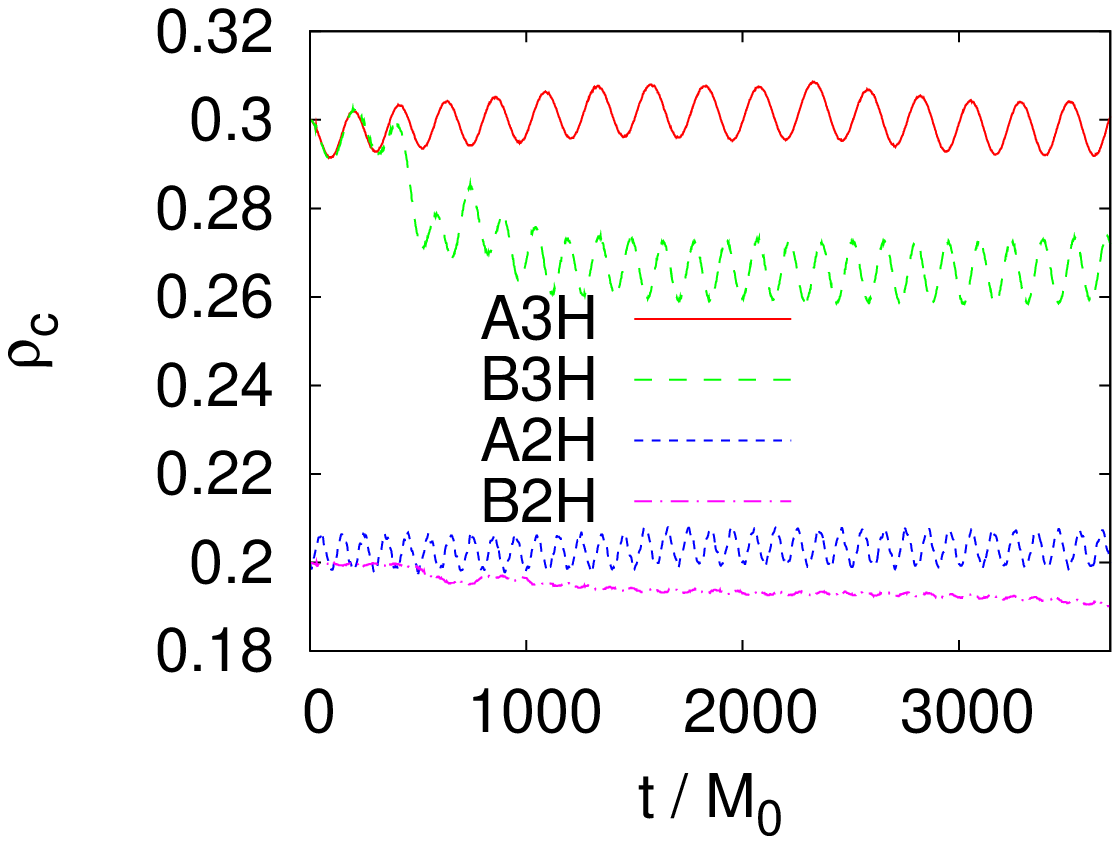}
\epsfxsize=3.in
\leavevmode
\epsffile{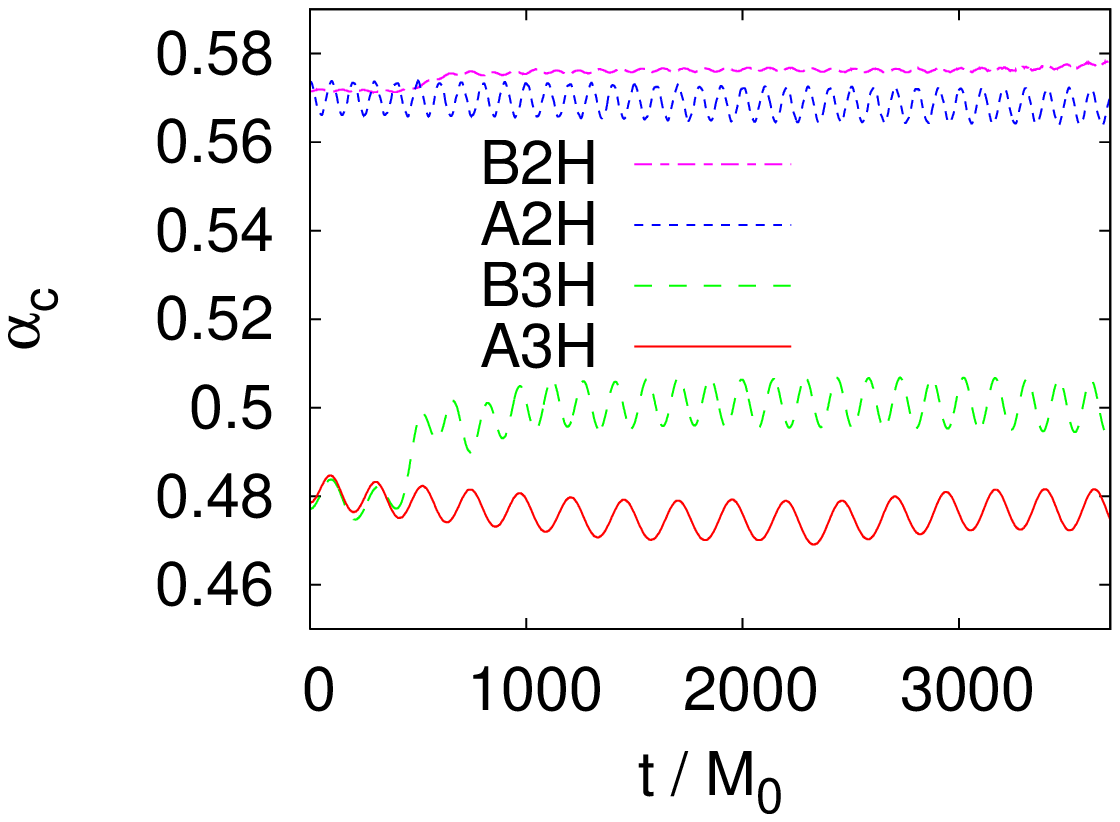}
\vspace{-2mm}
\caption{Evolution of central density and central value of the
  lapse function for models A3H (solid curves), B3H (long-dashed curves), 
  A2H (dashed curves), and B2H (dashed-dotted curves).  The
  units of time is initial ADM mass $M_0$.
\label{FIG1}}
\end{figure*}

We also prepare a variety of rotating neutron stars. In this paper, we
focus only on rigidly rotating neutron stars with a moderate
compactness; $\rho_c$ is fixed to be 0.200.  For neutron stars with
$\Gamma=2$ polytrope, the maximum ratio of rotational kinetic energy
to gravitational potential energy, $T/W$, is $\sim 0.09$ for compact
neutron stars with $R/M \approx 6$ \cite{CST}.  Here, at the maximum
ratio, velocity at the equatorial surface of the star is equal to the
Kepler velocity.  Taking into account this fact, we prepare rotating
stars with $0 < T/W \alt 0.09$. Specifically, we consider the
following four sequences for studying dependence of stability on the
rotation rate. First we consider rapidly rotating stars with
$T/W=0.08$--0.09 or $T/W \approx 0.06$ and with $0.001 \alt E_{\rm 
EM}/W \alt 0.02$.  The models in these categories are specified with
models ``R***'' and ``M***'' (see the caption of Table I for the
meaning of ***). In the third and fourth sequences, we approximately
fix the values of $E_{\rm EM}/W$ as 0.04 and 0.01 but change the
values of $T/W$ for a wide range.  The models in these categories are
specified with model ``E***'' and ``S***''.  By studying stability of
these models, the dependence of stability criterion on $T/W$ and
$E_{\rm EM}/W$ is clarified.

\section{Numerical simulation}

\subsection{Choosing the grid points and atmosphere}

Numerical simulation was performed assuming the axial symmetry 
as well as the equatorial ($z=0$) plane symmetry. For covering 
computational domain, a nonuniform grid 
of the following grid structure is adopted for $\varpi$ and $z$: 
\beqn
x^k(i)=\left\{
\begin{array}{ll}
i\Delta x & 1 \leq i \leq N_0 \\
i\Delta x + \xi \Delta i \Delta x& ~ \\
~\times \log [ \cosh \{(i-N_0)/\Delta i \}]  &  N_0 < i \leq N.
\end{array}
\right.
\eeqn
Here, $x^k$ denotes $\varpi$ or $z$, $\Delta x$ is the grid spacing in
the inner region, and $N_0$, $N$, $\Delta i$, and $\xi$ are constants
which determine the grid structure of the outer part. In this grid
setting, the inner domain with $0 \leq \varpi \leq N_0 \Delta x$ and
$0 \leq z \leq N_0 \Delta x$ is covered by a uniform grid.  Neutron
stars are always covered in an inner region with $r < 2N_0 \Delta 
x/3$.  $N_0$ is chosen to be 150 in this paper. 

In the present simulation, mass ejection and expansion of stars do not
occur in a remarkable manner, and hence, it is not necessary to
resolve the outer region as accurately as the inner region where
neutron stars are located. Thus, we prepare a rather large grid
spacing for the outer region choosing $\Delta i=50$ and $\xi=10$. $N$
is set to be 240.  In the following, we present results with this grid
setting for all the cases.  With this setting, outer boundaries along
each axis are located at $L \approx 800\Delta x$ which is
approximately equal to eight stellar radii ($\approx 8R$).  These are
large enough for excluding spurious effects from outer boundaries at
least for $t \alt 3000M_0$. Indeed, we performed a simulation for
$N=220$ (i.e., $L \approx 600\Delta x$) while fixing other parameters
for the grid structure, and found that results depend very weakly on
$N$ (i.e., $L$; see Fig. \ref{FIG9}). Nevertheless, for smaller values
of $L$, the spurious effects coming from the outer boundaries are
serious: For $L=300 \Delta x \approx 3R$ and $600 \Delta x \approx
6R$, the computation crashes eventually at $t \sim 2000M_0$ and
$2700M_0$, respectively, in the chosen non uniform grid. However, note
that the time at which computation crashes depends on the grid
structure and for the uniform-grid case, the computation does not
crash at $t=3500 M_0$ even for $L=3R$.

For a convergence test, we performed additional simulations for model
B3H, choosing uniform grid with $N=240$, 300, and 360.  For each case,
the equatorial radius of the neutron star is covered by 80, 100, and
120 grid points, respectively. Thus, for $N=300$, the grid resolution
for the inner region with $i \leq 150$ is the same as that for the
nonuniform grid. In these uniform grids, outer boundaries are located
at three stellar radii ($L=3R$), but in these cases, the computation
does not crash until $t=3500M_0$.  A simulation was also performed in
the nonuniform grid with $N=220$, in which $L \approx 600\Delta x$, as
mentioned above.  Comparison of the results for five simulations
indicates that the resolution of our typical grid setting and location
of outer boundaries are fine enough to derive quantitative results
within 20--30\% error (see the last two paragraphs of
Sec. \ref{secnonrot}).

Because any conservation scheme of hydrodynamics is unable to evolve a
vacuum, we have to introduce an artificial atmosphere outside neutron
stars.  We initially assign a small rest-mass density of magnitude
$\rho_{\rm at}=\rho_{\rm max} \times 10^{-8}$ where $\rho_{\rm max}$
is the maximum rest-mass density of the neutron star. With such
choice, the total amount of the rest mass of the atmosphere is about
$10^{-5}$ of the rest mass of the neutron star. Thus, the accretion of
the atmosphere onto the neutron star plays a negligible role for their
evolution.

\subsection{Numerical results} \label{sec:results}

\subsubsection{Nonrotating case}\label{secnonrot}

\begin{figure}[t]
\epsfxsize=2.2in
\leavevmode
\hspace{-1.cm}\epsffile{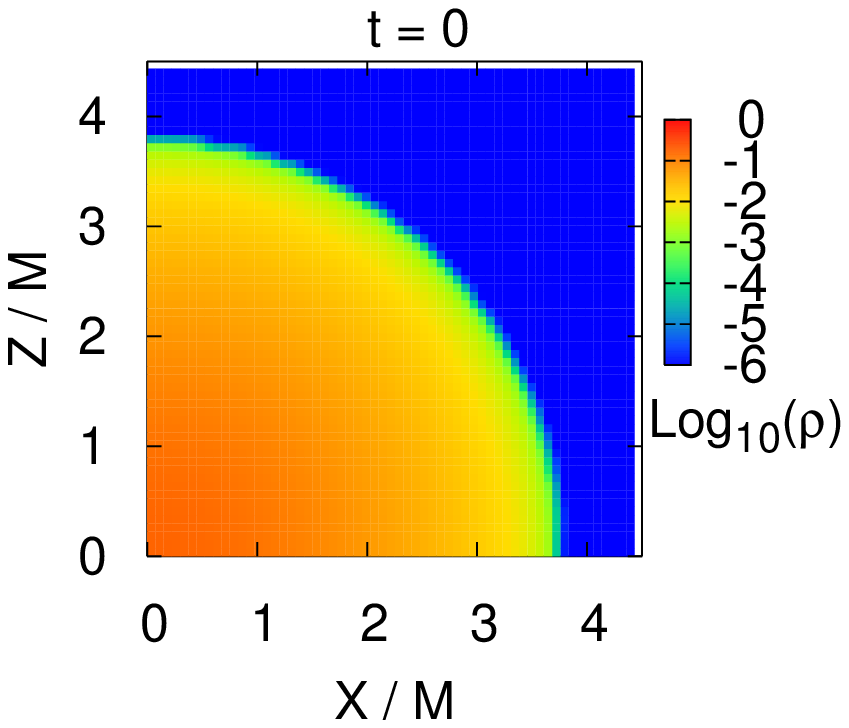}
\epsfxsize=2.2in
\leavevmode
\hspace{-1.7cm}\epsffile{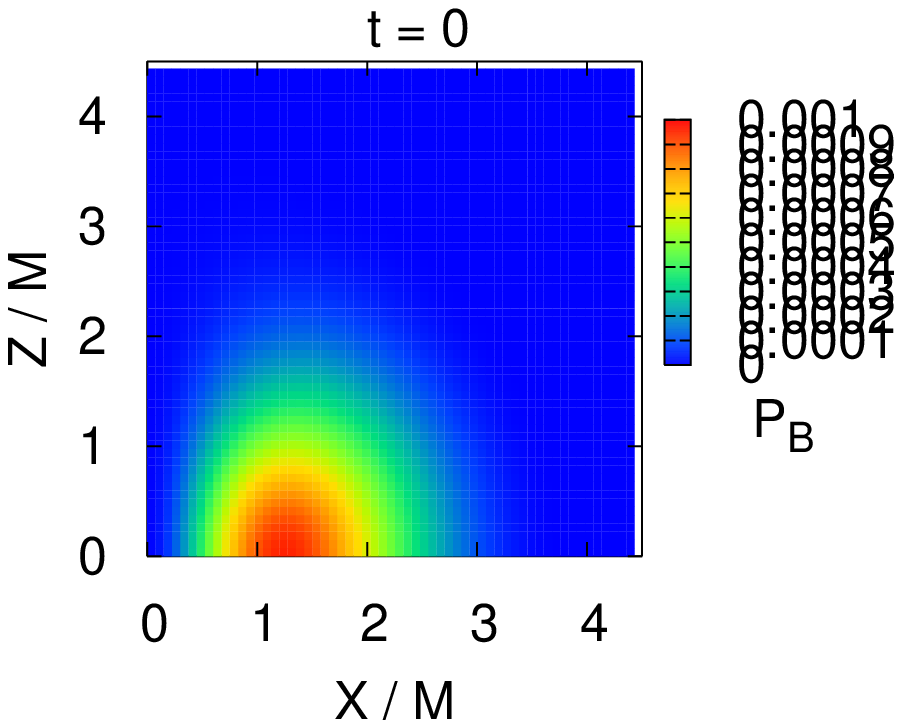} \\
\epsfxsize=2.2in
\leavevmode
\hspace{-1.cm}\epsffile{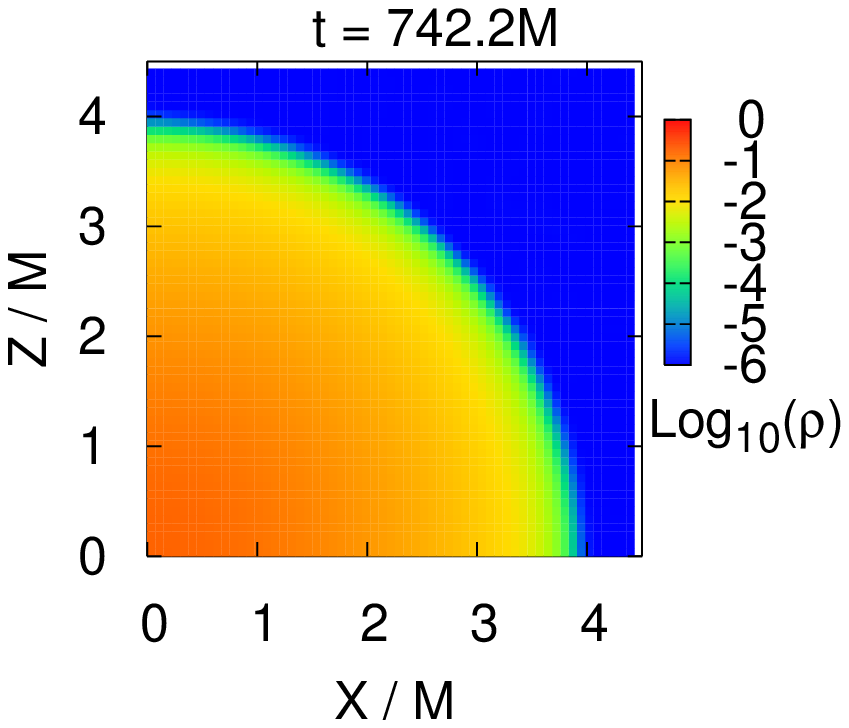}
\epsfxsize=2.2in
\leavevmode
\hspace{-1.7cm}\epsffile{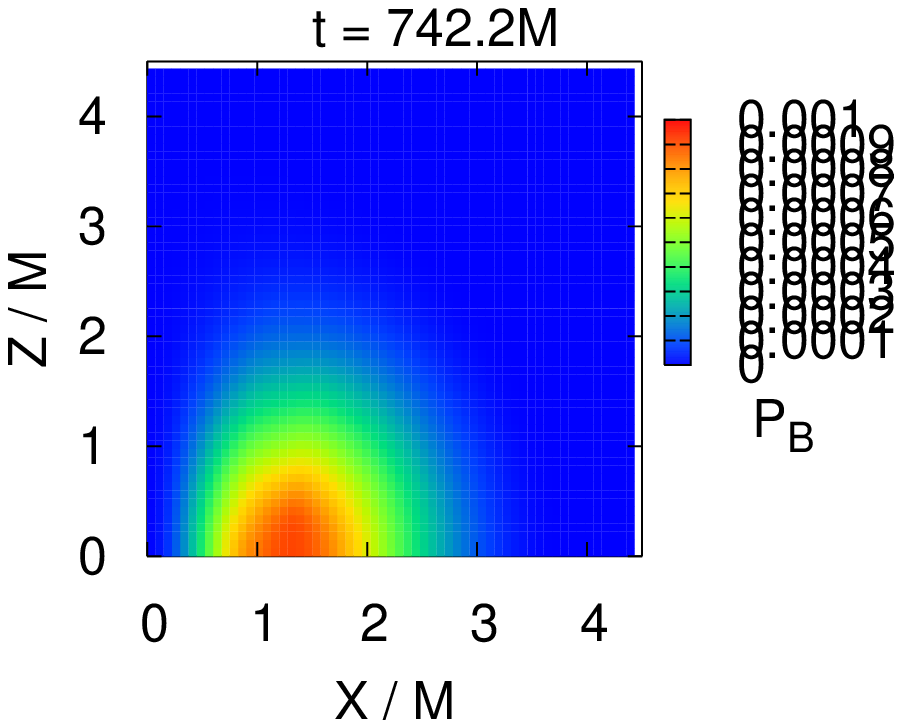}
\vspace{-0.2mm}
\caption{Snapshots for profiles of rest-mass density and 
  magnetic pressure, $b^2/2$, for model A3H which is dynamical stable. 
\label{FIG2}}
\end{figure}

\begin{figure*}[t]
\epsfxsize=2.2in
\leavevmode
\hspace{-0.5cm}\epsffile{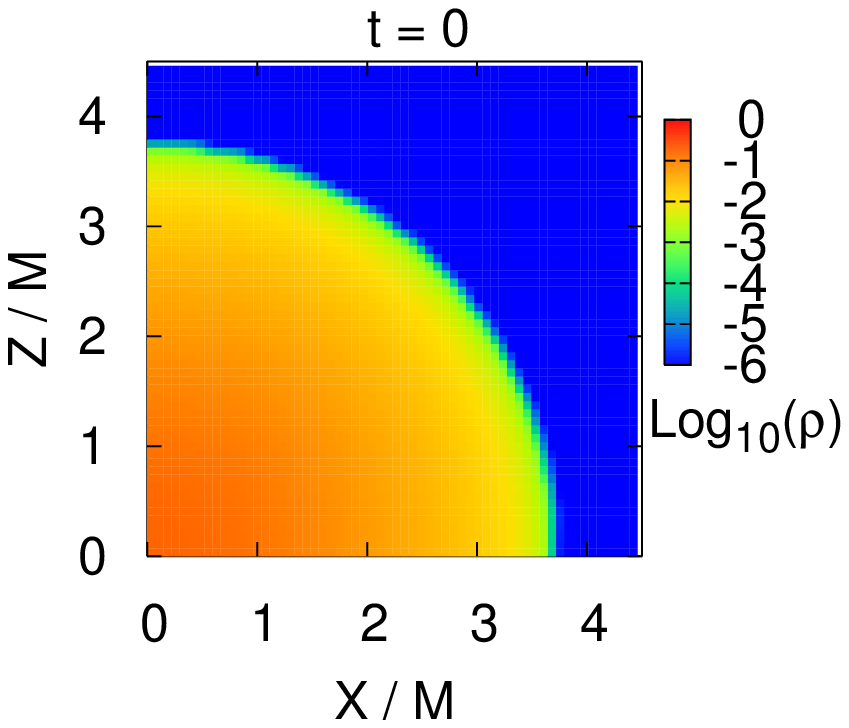}
\epsfxsize=2.2in
\leavevmode
\hspace{-1.7cm}\epsffile{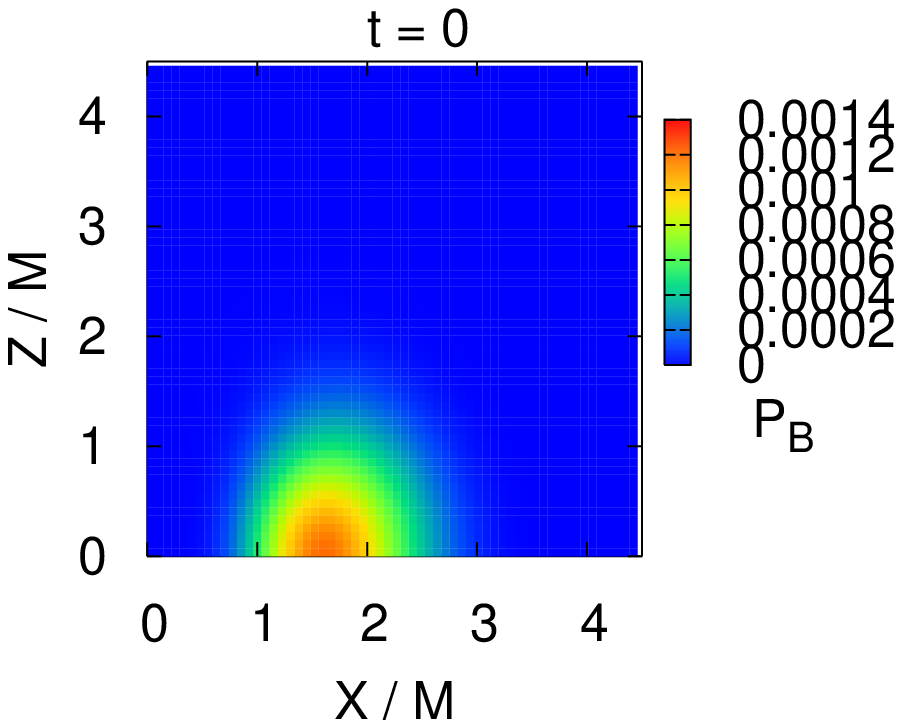}
\epsfxsize=2.2in
\leavevmode
\hspace{-1.4cm}\epsffile{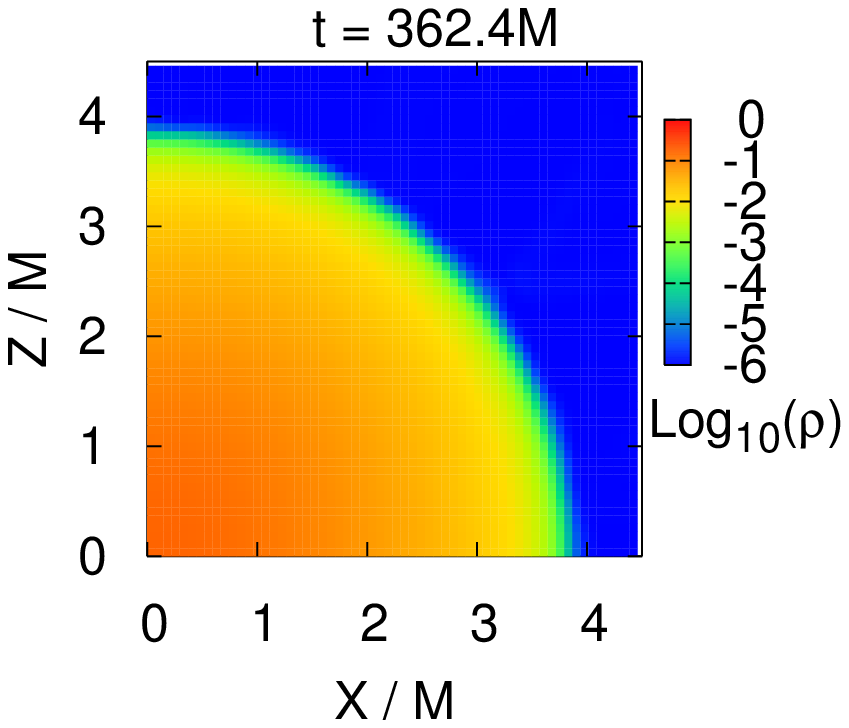}
\epsfxsize=2.2in
\leavevmode
\hspace{-1.7cm}\epsffile{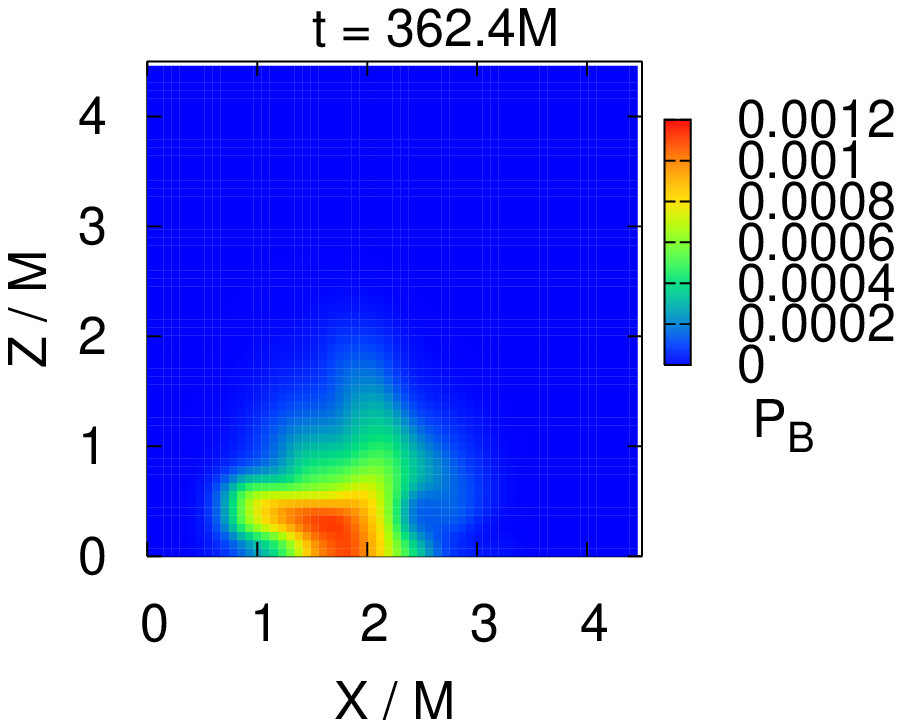}\\
\vspace{-0.5cm}
\epsfxsize=2.2in
\leavevmode
\hspace{-0.5cm}\epsffile{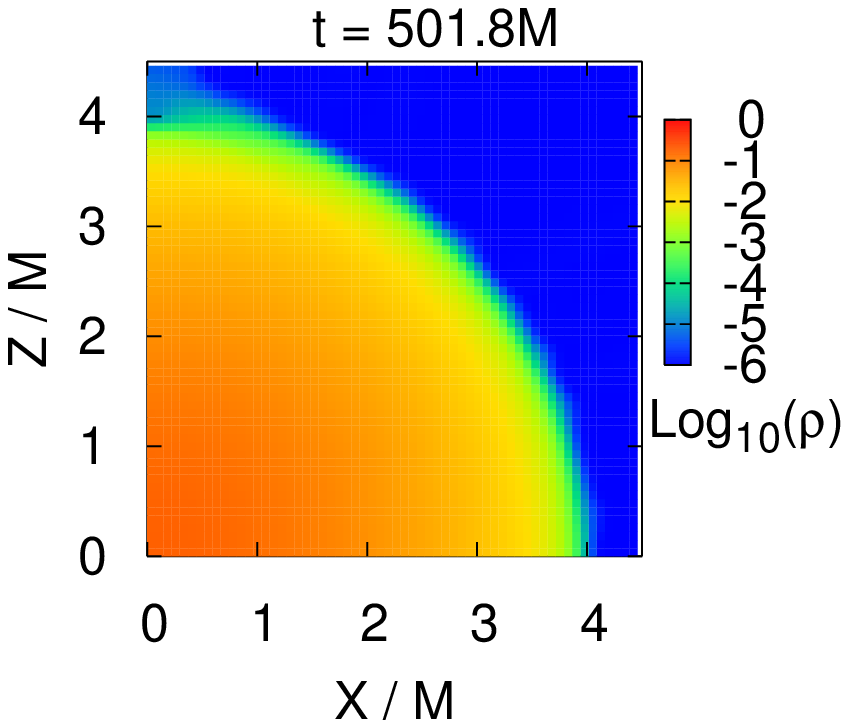}
\epsfxsize=2.2in
\leavevmode
\hspace{-1.7cm}\epsffile{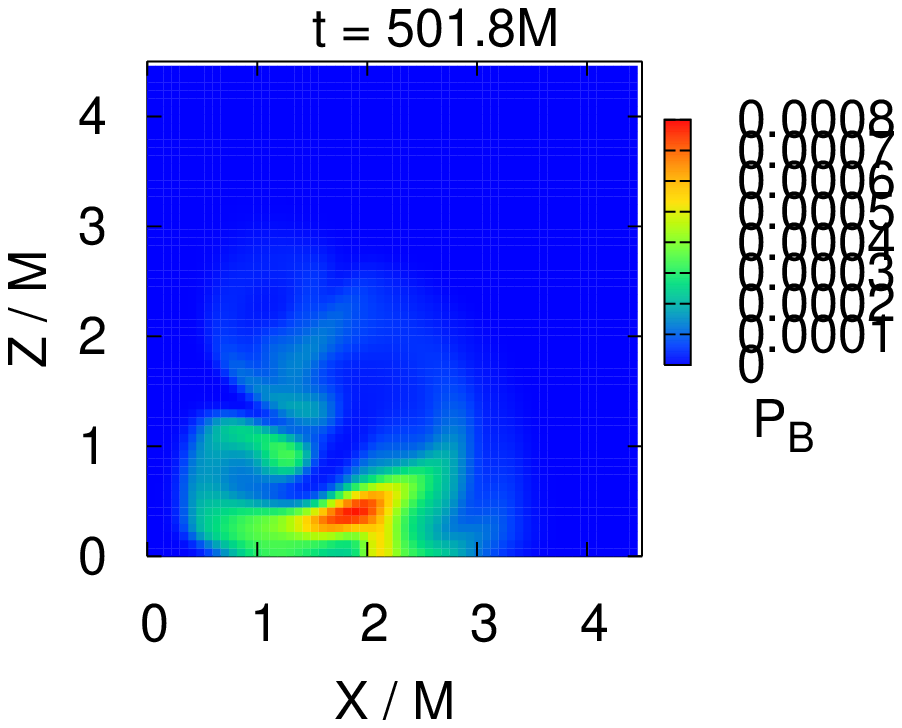}
\epsfxsize=2.2in
\leavevmode
\hspace{-1.4cm}\epsffile{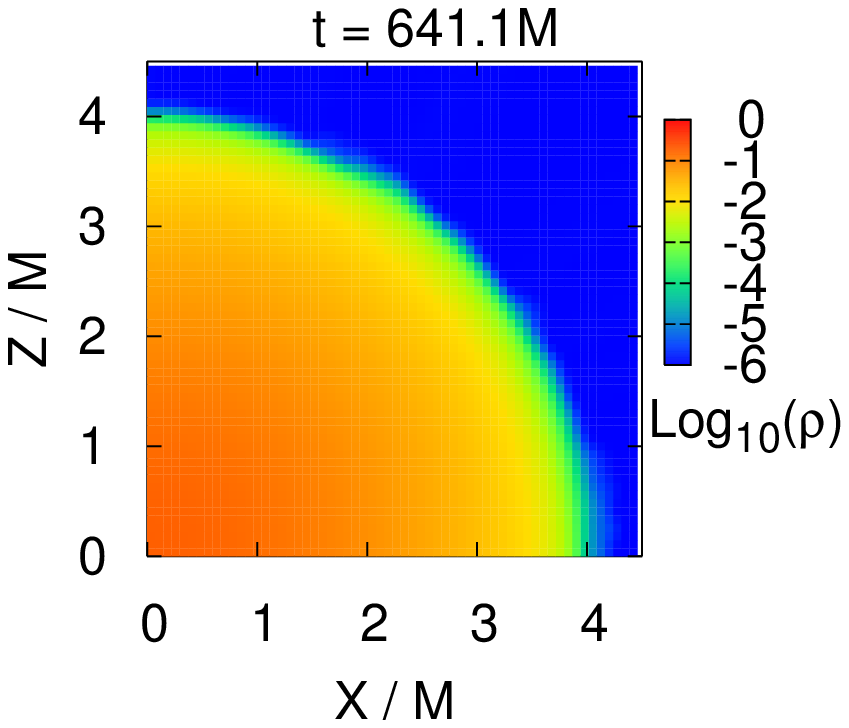}
\epsfxsize=2.2in
\leavevmode
\hspace{-1.7cm}\epsffile{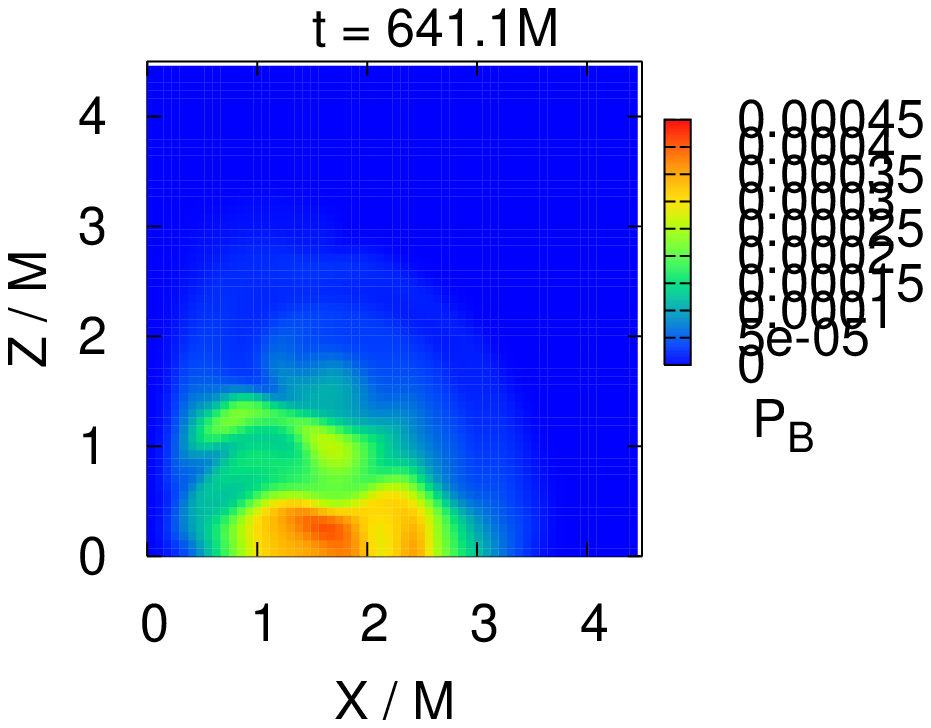}\\
\vspace{-0.5cm}
\epsfxsize=2.2in
\leavevmode
\hspace{-0.5cm}\epsffile{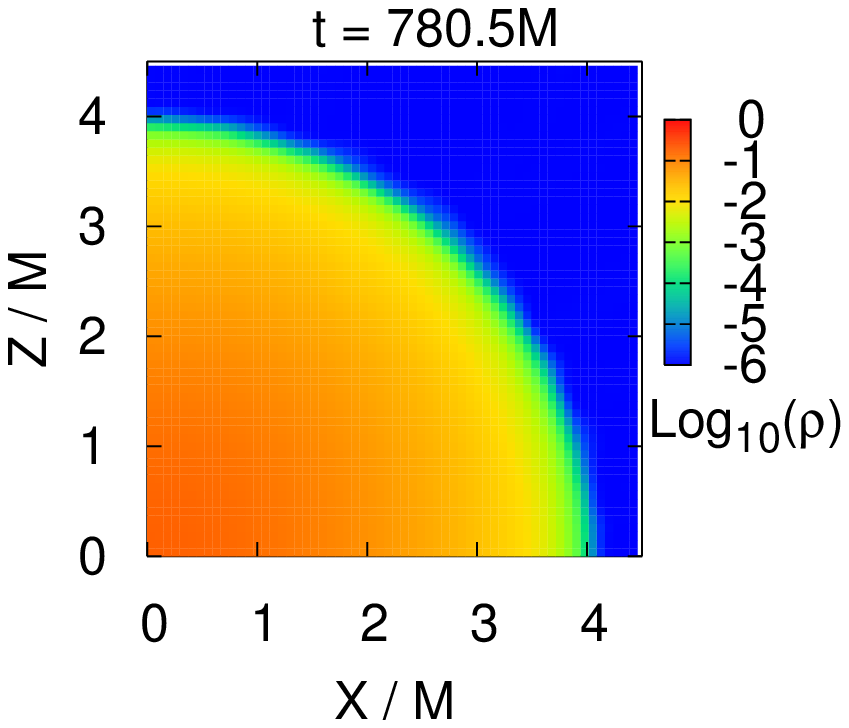}
\epsfxsize=2.2in
\leavevmode
\hspace{-1.7cm}\epsffile{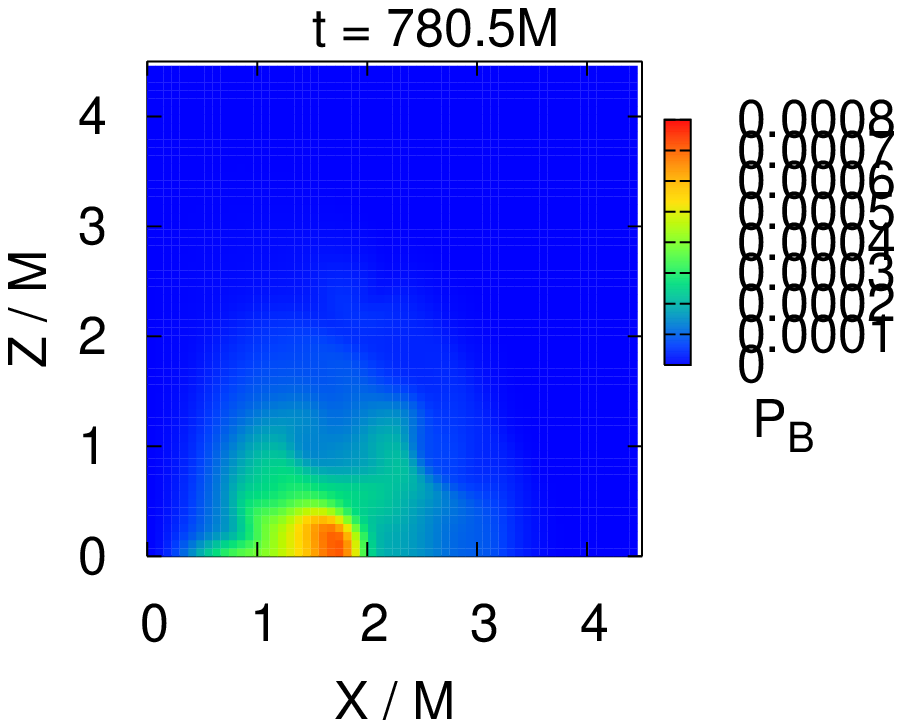}
\epsfxsize=2.2in
\leavevmode
\hspace{-1.4cm}\epsffile{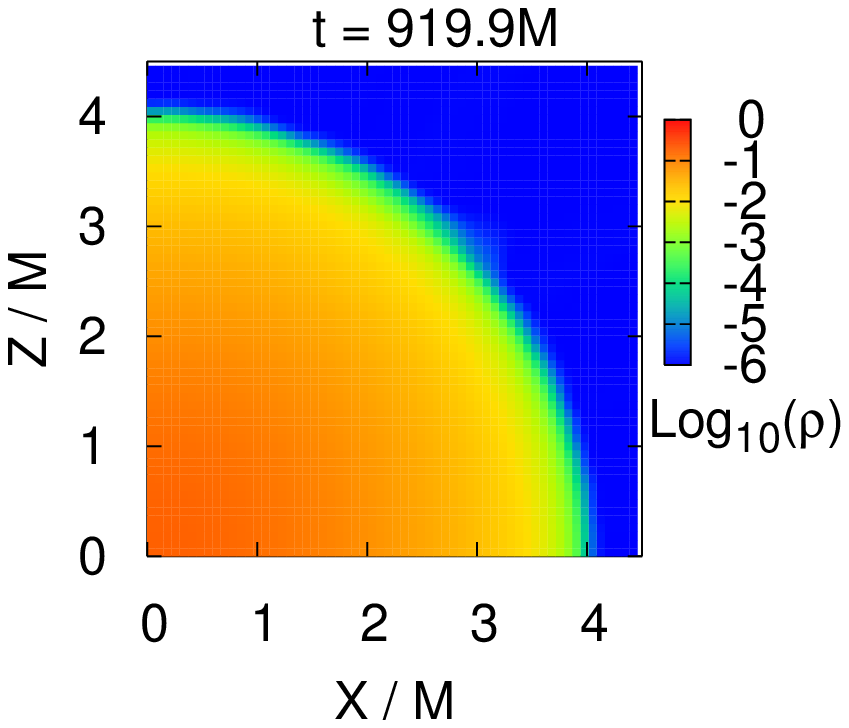}
\epsfxsize=2.2in
\leavevmode
\hspace{-1.7cm}\epsffile{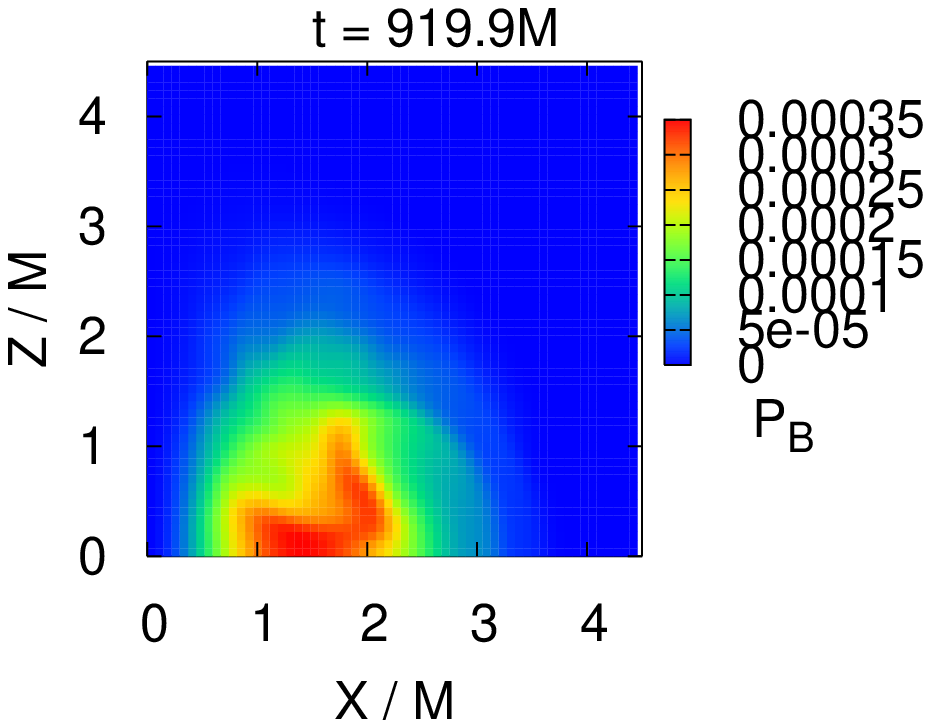}\\
\vspace{-0.5cm}
\epsfxsize=2.2in
\leavevmode
\hspace{-0.5cm}\epsffile{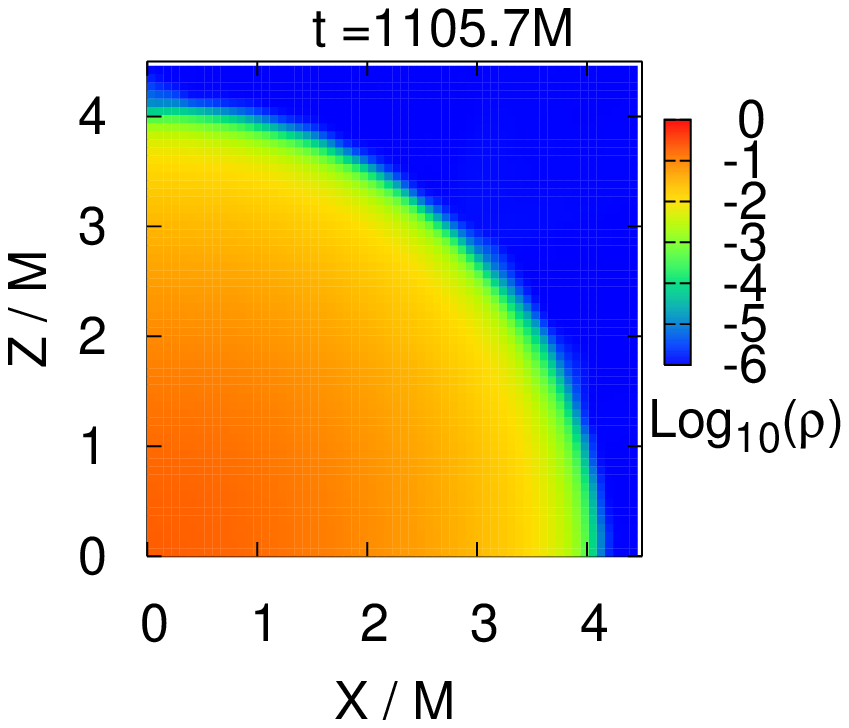}
\epsfxsize=2.2in
\leavevmode
\hspace{-1.7cm}\epsffile{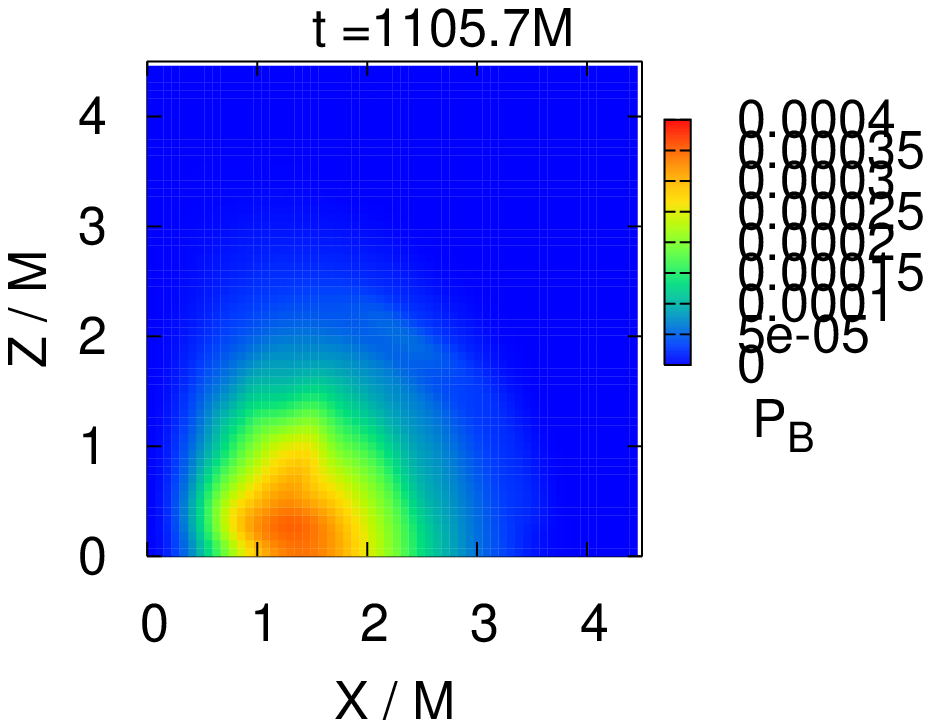}
\epsfxsize=2.2in
\leavevmode
\hspace{-1.4cm}\epsffile{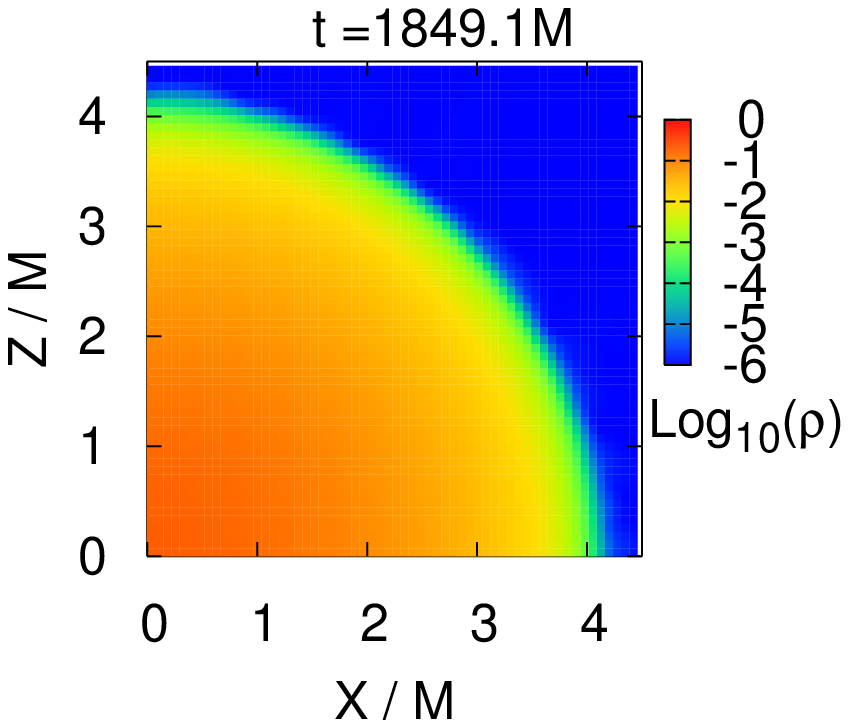}
\epsfxsize=2.2in
\leavevmode
\hspace{-1.7cm}\epsffile{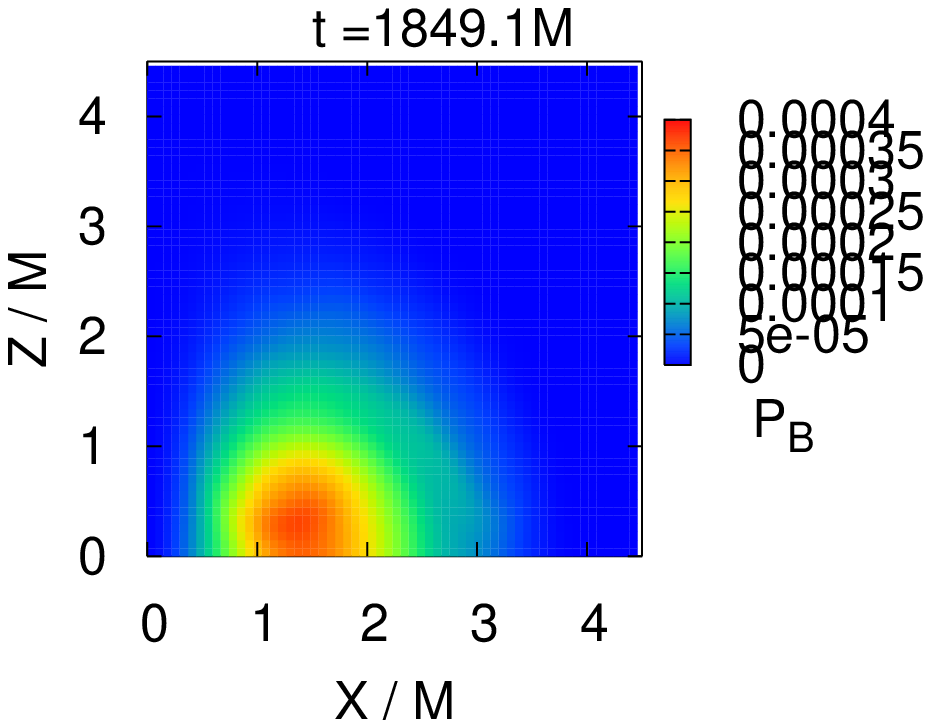}
\vspace{-2mm}
\caption{Snapshots for profiles of rest-mass density and 
  magnetic pressure for model B3H. 
\label{FIG3}}
\end{figure*}

\begin{figure}[th]
\epsfxsize=3.4in
\leavevmode
\epsffile{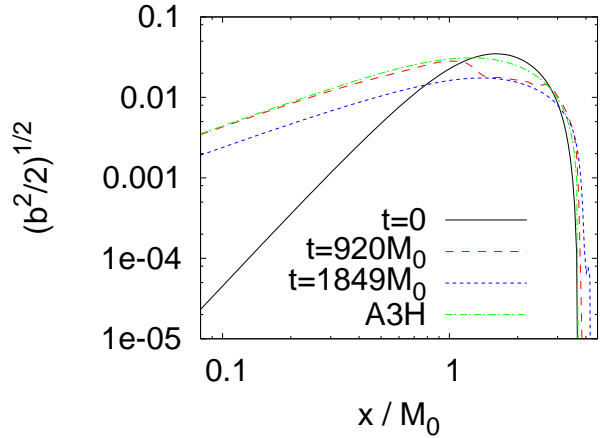}
\vspace{-2mm}
\caption{Profile of square root of magnetic pressure,
  $\sqrt{b^2/2}$, along $\varpi$-axis at selected
  time slices, $t/M_0=0$, 920, and 1849.1 for model B3H.  For
  comparison, the profile for model A3H at $t=0$ is shown together. 
\label{FIG4}}
\end{figure}

\begin{figure*}[b]
\epsfxsize=1.8in
\leavevmode
\epsffile{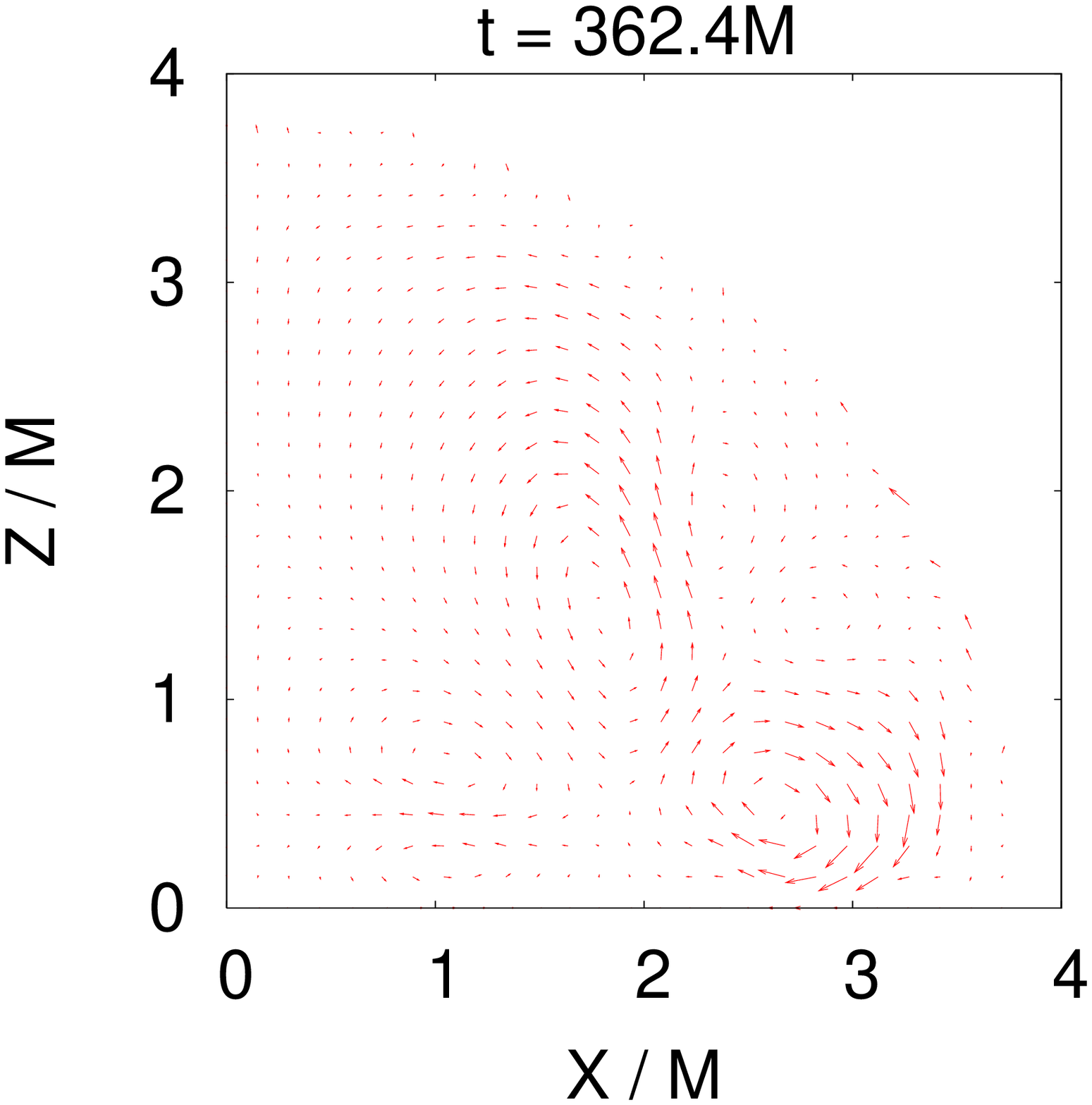}
\epsfxsize=1.8in
\leavevmode
\hspace{-0.2cm}\epsffile{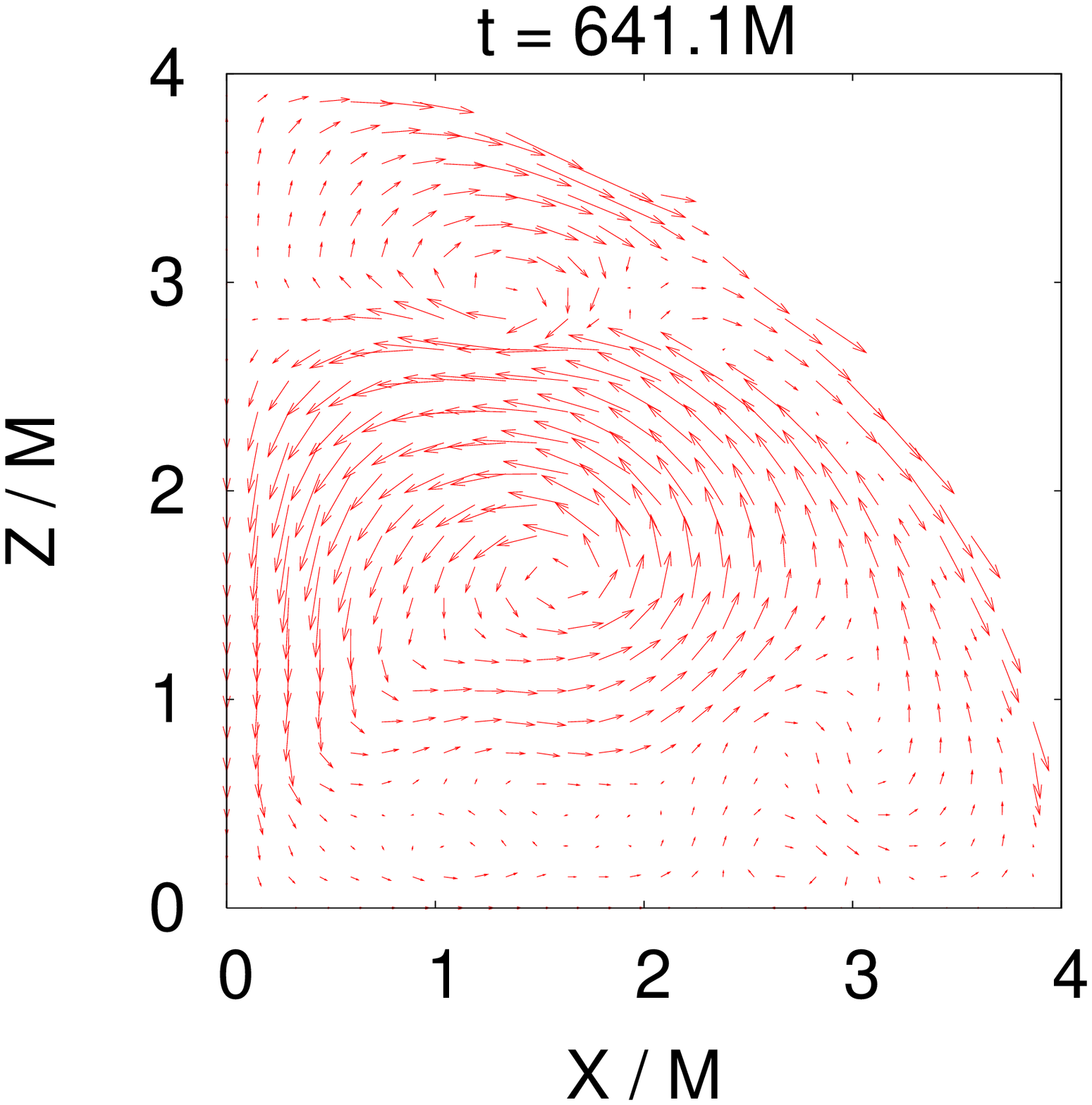}
\epsfxsize=1.8in
\leavevmode
\hspace{-0.2cm}\epsffile{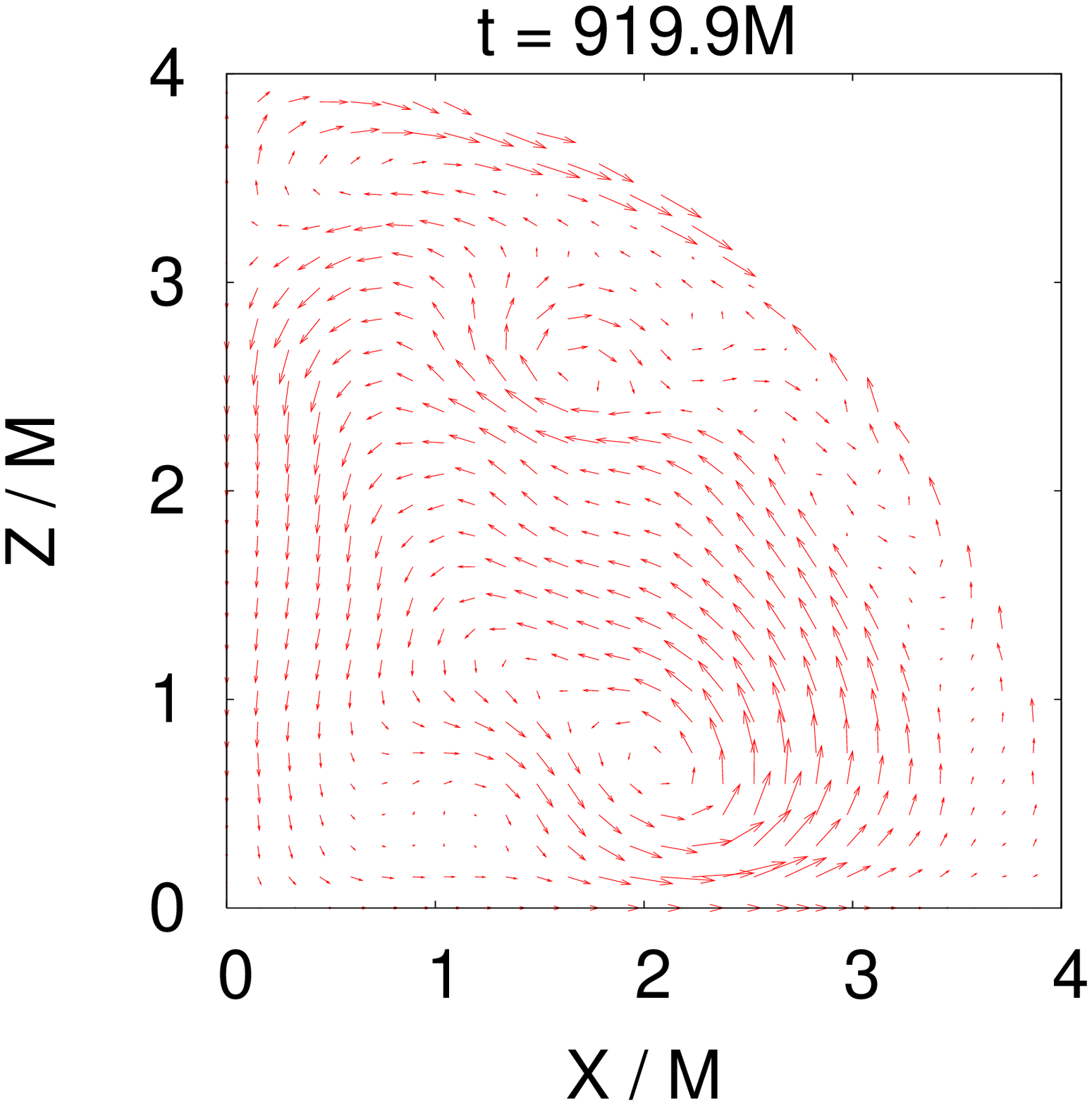}
\epsfxsize=1.8in
\leavevmode
\hspace{-0.2cm}\epsffile{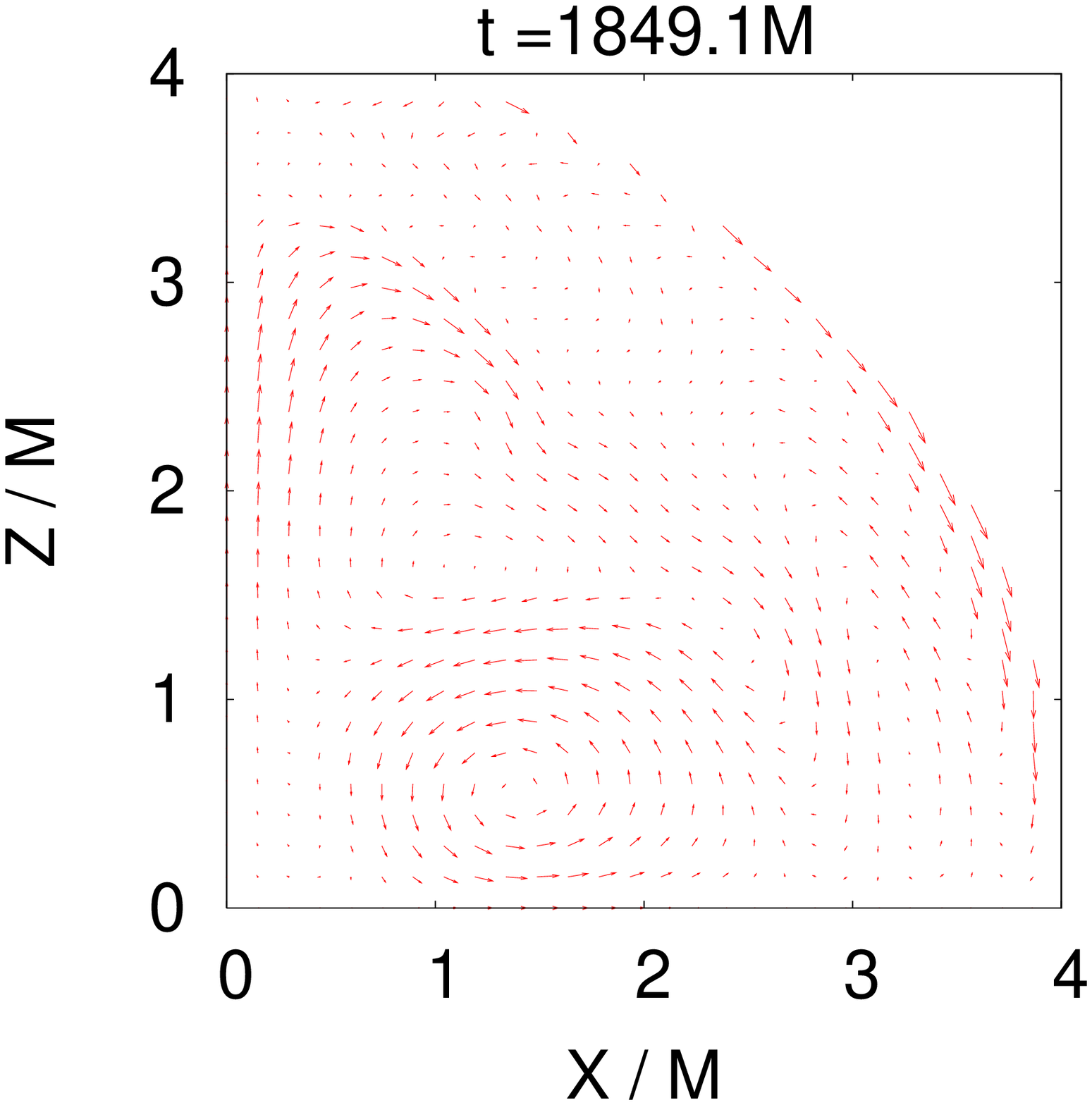}
\caption{Snapshots of velocity fields for model B3H at selected
  time slices, $t/M_0=362.4$, 641.1, 919.9, and 1849.1
  (cf. Fig. \ref{FIG3} for the magnetic field profile at the same
  time slices).  For each time, the maximum velocity (the maximum
  value of $\sqrt{(v^{\varpi})^2+(v^z)^2}$) is 0.039, 0.047, 0.041,
  and 0.037, respectively.
\label{FIG5}}
\end{figure*}

We performed numerical simulations for all the models listed in Table
I.  All the simulations stably proceeded for a sufficiently long time
to more than $3000M_0$. For the case that unstable modes of MHD grow
in the Alfv\'en time scale, the simulation time is long enough to
determine the stability of each neutron star and to follow subsequent
evolution after the onset of the instability for our chosen models.
 
By the numerical simulations, we find that all the models with $k=1$
are stable and the profiles of density and magnetic field do not
change significantly during evolution. By contrast, all the
nonrotating models with $k \geq 2$ are unstable irrespective of the
magnetic field strength. In this case, the magnetic fields are
redistributed, and as a result, a convective motion is excited inside
the neutron star. The unstable stars slightly expand and the central
density decreases (cf. Fig. \ref{FIG1}).  Eventually, the star relaxes
to a new state which is stable against axisymmetric
perturbations. These conclusions hold irrespective of the magnetic
field strength and compactness of the neutron stars. In the following,
we describe characteristic features for stable and unstable stars
showing numerical results for specific models.

Figure \ref{FIG1} plots evolution of central density and central value
of the lapse function for models A3H, B3H, A2H, and B2H.  This
illustrates that for models A3H and A2H, the neutron stars simply
oscillate around their hypothetical equilibrium states.  The
oscillation is excited because the initial model deviates slightly
from the true equilibrium. The oscillation amplitude is larger for
$\rho_c=0.3$ than that for $\rho_c=0.2$. This reflects a fact that the
neutron stars with $\rho_c=0.3$ are close to the marginally stable
point against gravitational collapse, and a small perturbation induces
a large deviation from the equilibrium state.  In contrast to models
A3H and A2H, for models B3H and B2H, the central density and lapse
quickly change at $t \sim 500M_0$, implying that the stars
expand. This is due to the fact that the profile of magnetic fields is
modified during the evolution.

Figures \ref{FIG2} and \ref{FIG3} display snapshots for the profiles
of density and magnetic pressure for models A3H and B3H at selected
time slices.  For model A3H in which the neutron star is stable, the
profiles remain approximately static besides a slight oscillation. By
contrast, model B3H is dynamically unstable against redistribution of
magnetic fields: For $t \alt 400M_0$, the profile of the magnetic
pressure distribution gradually varies, and then, for $400 \alt t/M_0
\alt 1000$, the magnetic fields are redistributed violently.  As
described in Appendix A, unstable modes grow near the equatorial
plane as well as in a high latitude of $z \sim M_0$ and $\varpi \sim
2M_0$.  For $t \agt 1000M_0$, the profile approaches to a new state
which is stable against axisymmetric perturbation.  The maximum
magnetic pressure decreases after the onset of the instability. As a
result, the pinching effect by the toroidal magnetic fields are
weaken. This is the reason that the star slightly expands and the
central density decreases.

Figure \ref{FIG4} plots square root of magnetic pressure,
$\sqrt{b^2/2}$, along cylindrical axis ($\varpi$ axis) at selected
time slices ($t/M_0=0$, 920, and 1849.1) for model B3H.  For comparison,
the profile for model A3H at $t=0$ is shown together. For model B3H in
which $k=2$, the magnetic field strength initially distributes
approximately in proportional to $\varpi^3$ for $\varpi \alt M_0$. As
a result of the growth of an instability, this profile
changes, and eventually, the magnetic field strength becomes
approximately proportional to $\varpi$ for $\varpi \alt M_0$. 
Indeed, in the relaxed state,
the profile is similar to that for model A3H in which
$k=1$ and $\sqrt{b^2/2} \propto \varpi$ for $\varpi < M_0$.  This
indicates that magnetized stars with $k=1$ are attractors for the
unstable star in axial symmetry.

Figure \ref{FIG5} plots velocity vector fields $(v^{\varpi}, v^z)$
at selected time slices for model B3H. As shown in this figure,
convective motion and circulation are excited during the nonlinear
evolution after the onset of instability.  The velocity of the
convective motion becomes maximum during the nonlinear evolution of
the instability at $t \sim 600$--700$M_0$ and the maximum velocity of
this motion is $\sim 5\%$ of the speed of light. It is interesting to
point out that the convective motion is present even after the
magnetic field profile approximately relaxes to a stable state.

\begin{figure*}[thb]
\epsfxsize=3.in
\leavevmode
\epsffile{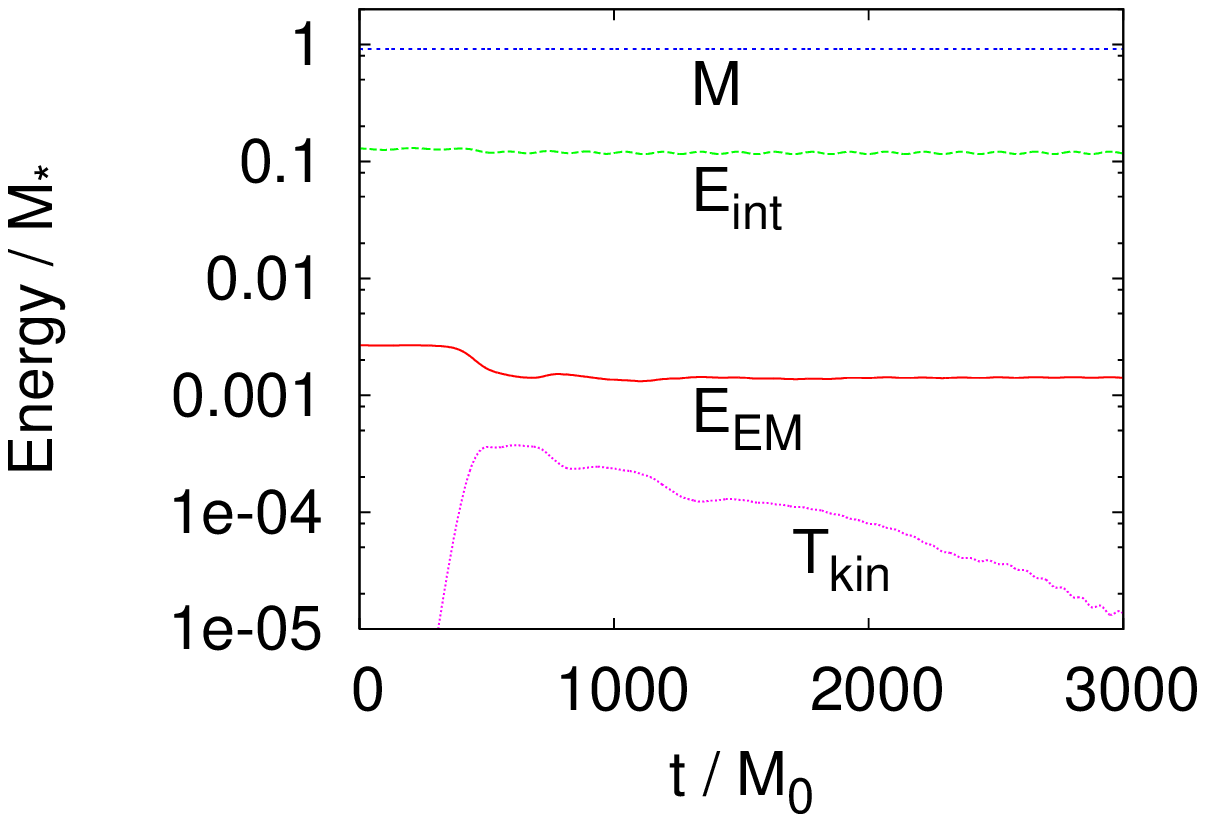}
\epsfxsize=3.in
\leavevmode
\epsffile{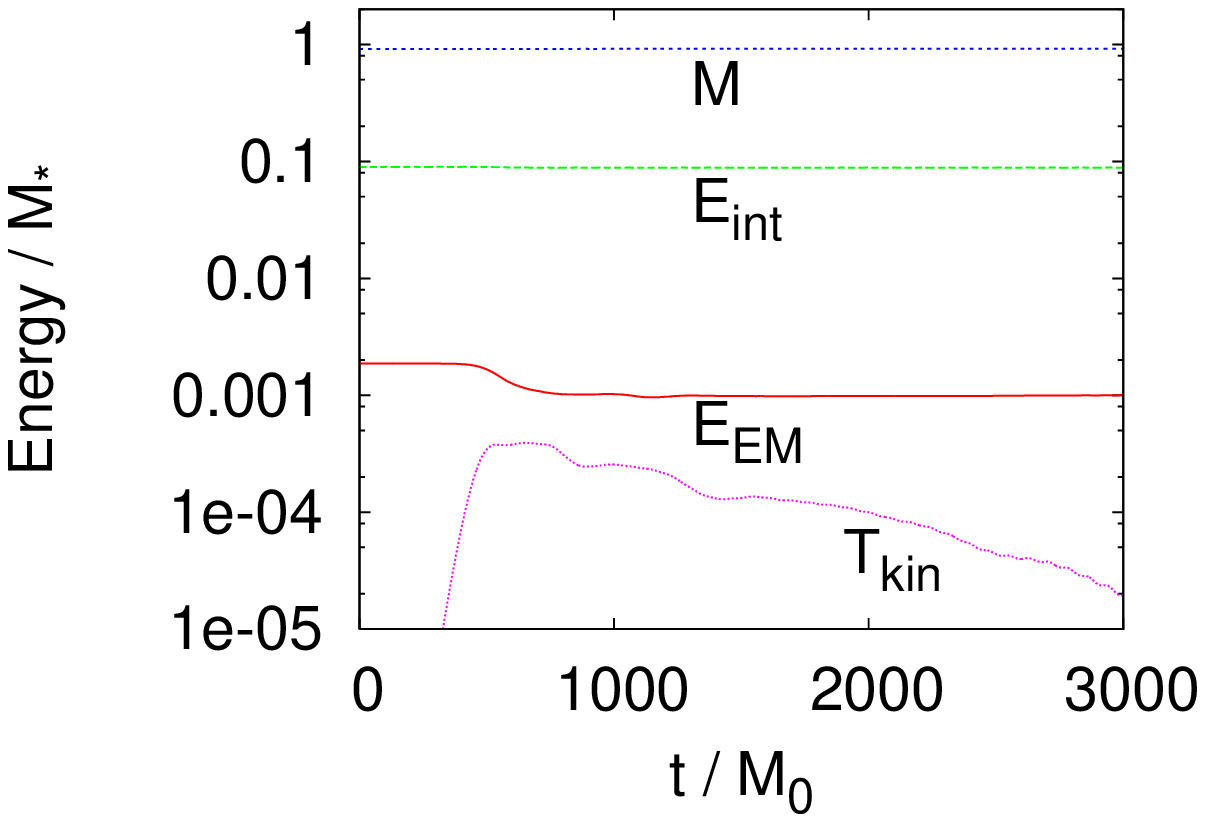}
\caption{Evolution of the ADM mass, internal thermal, kinetic, and
  electromagnetic energy for model B3H (left) and model B2H (right).
\label{FIG6}}
\end{figure*}

Figure \ref{FIG6} plots the evolution of the ADM mass, internal thermal,
kinetic, and electromagnetic energy for model B3H. This shows that the
ADM mass is approximately constant, and internal thermal energy does not vary
significantly. By contrast, the electromagnetic energy decreases
significantly after the onset of dynamical instability, and with the
quick decrease, the kinetic energy increases steeply. This is because
the convective motion is induced by the dynamical instability.  In
other words, the electromagnetic energy is transformed into the
kinetic energy.

The kinetic energy increases up to $\sim 30\%$ of the electromagnetic
energy for model B3H. (Possible error size of the kinetic energy is
20--30\% as we discussed later; cf. Fig. 9.)  This holds for models
with $\rho_c=0.3$ and $k=2$.  For $\rho_c=0.2$ and $k=2$, $T_{\rm
  kin}$ increases to $\sim 0.5 E_{\rm EM}$ and for models with $k=3$,
$T_{\rm kin}$ increases to $\sim 0.6 E_{\rm EM}$. All these results 
indicate that the kinetic energy of the convective motion can reach to a
value approximately as large as the electromagnetic energy for the
case that the instability grows. 

This result suggests that for a protoneutron star with strong 
toroidal magnetic fields (see Sec. I for discussion), the kinetic 
energy of the convective motion may reach 
\beqn 
T_{\rm kin}
\sim 10^{50} \biggl({B^T \over 10^{16}~{\rm G}}\biggr)^2 
\biggl({R   \over 15~{\rm km}}\biggr)^3~{\rm ergs}.\label{conv} 
\eeqn 
Because the kinetic energy could reach to such a large value, the
convective motion triggered by this type of instability inside a
protoneutron star may affect supernova explosion. 

\begin{figure*}[t]
\epsfxsize=3.in \leavevmode \epsffile{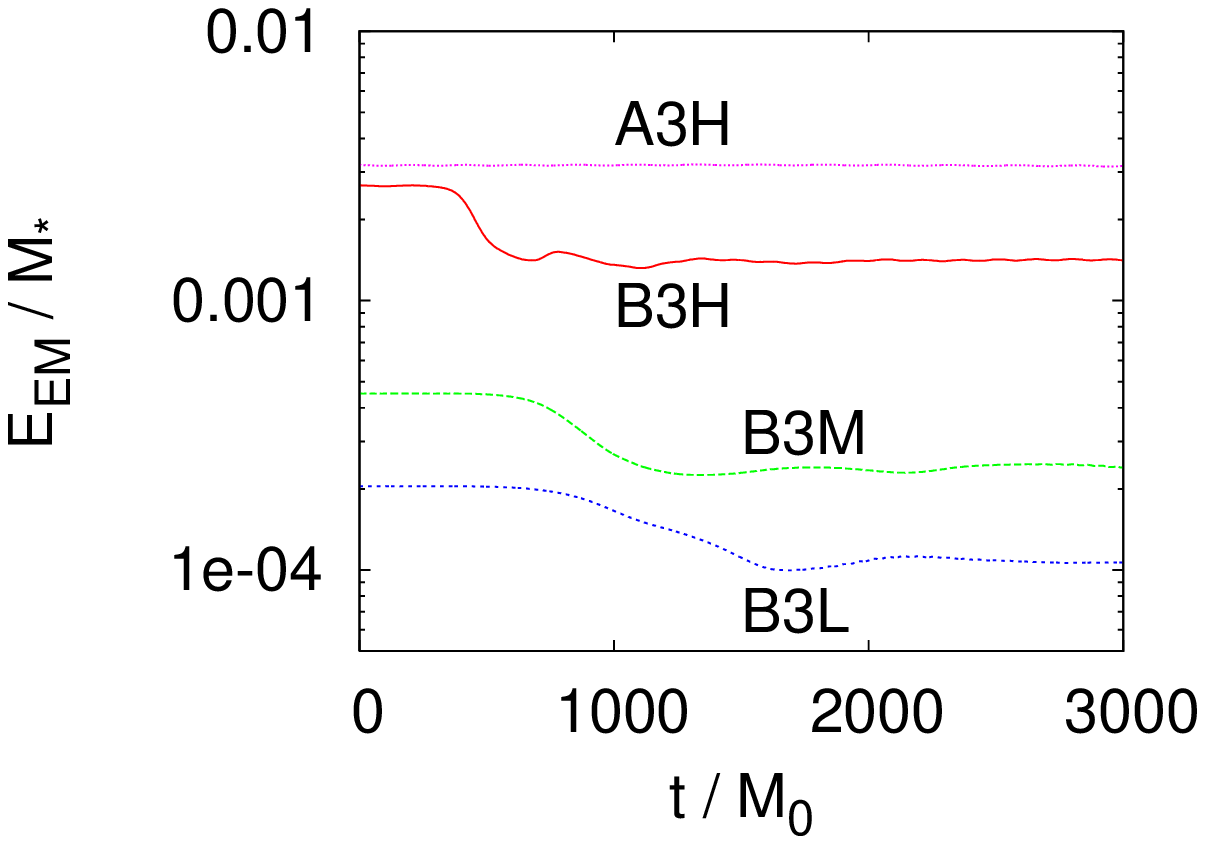} 
\epsfxsize=3.in \leavevmode \epsffile{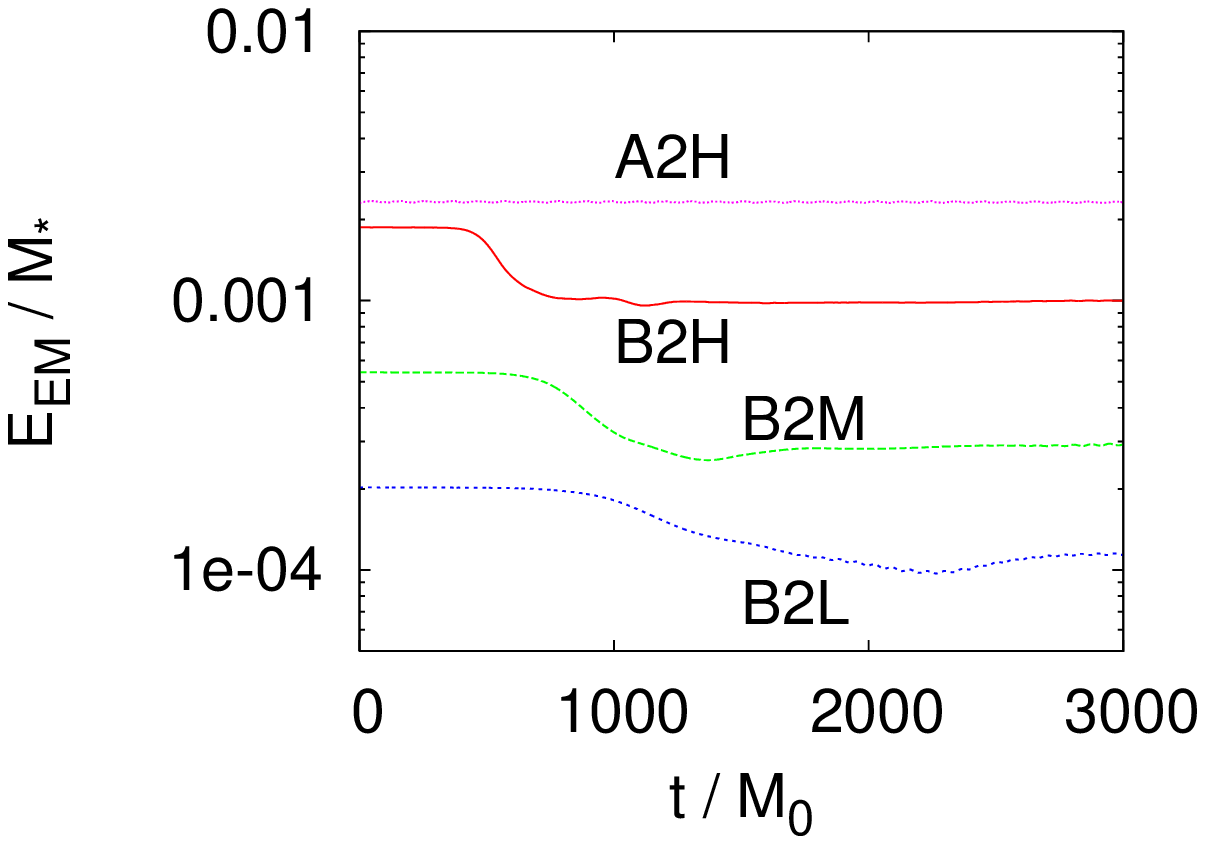}
\caption{Left: Evolution of electromagnetic energy for models B3H,
  B3M, B3L, and A3H.  Right: The same as the left panel but for models
  B2H, B2M, B2L, and A2H.
\label{FIG7}}
\end{figure*}

\begin{figure*}[t]
\epsfxsize=3.in 
\leavevmode 
\epsffile{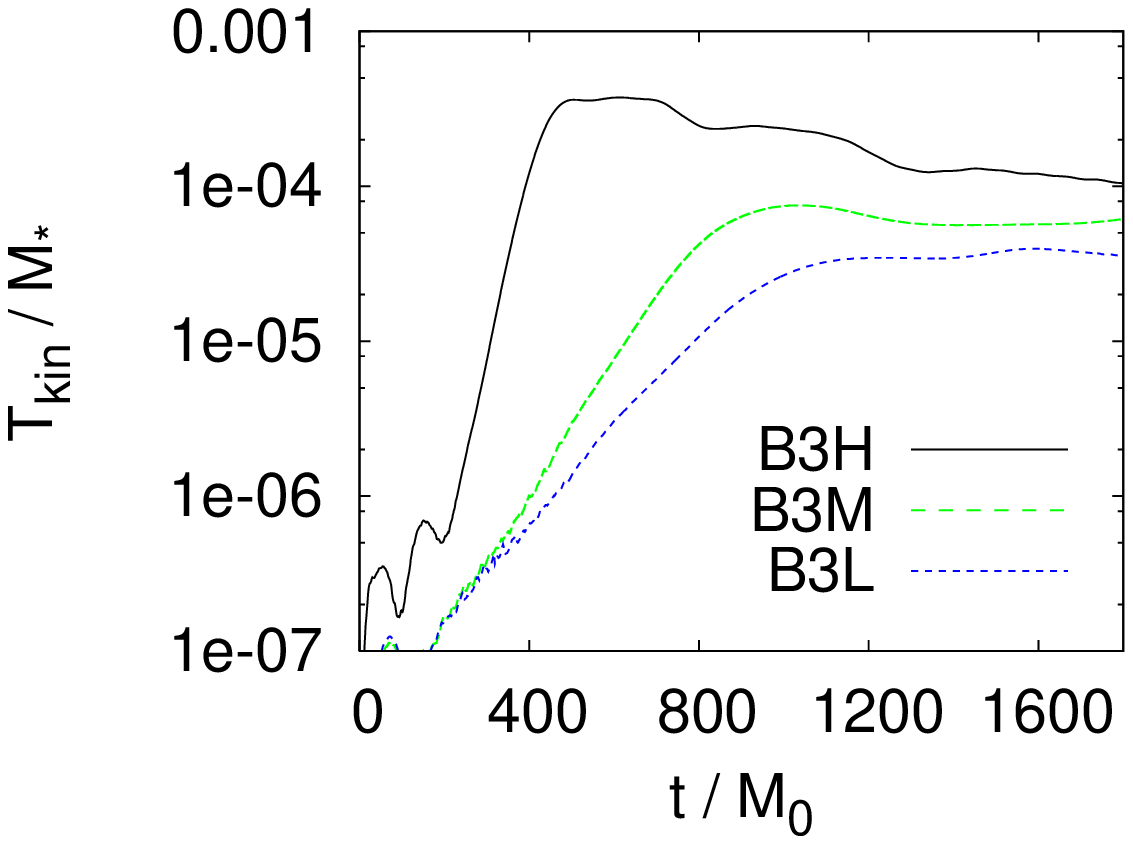}
\epsfxsize=3.in 
\leavevmode 
\epsffile{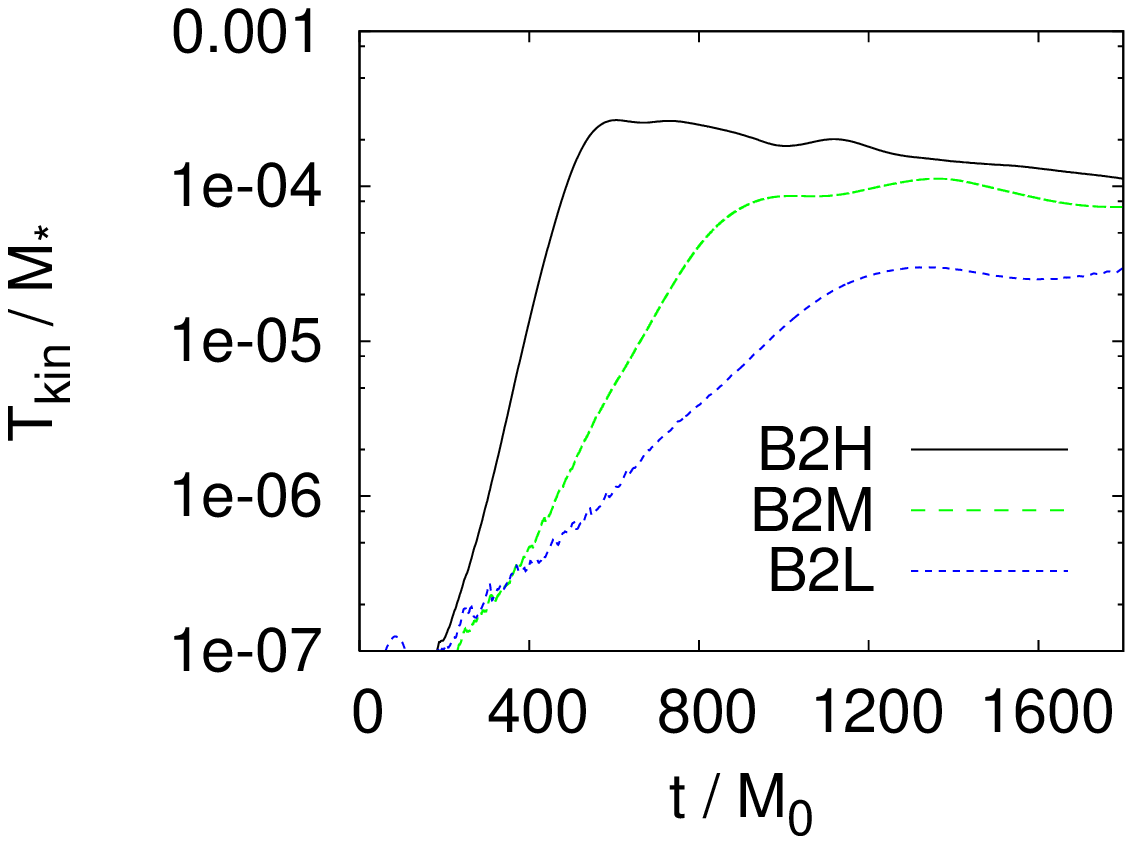}
\caption{$T_{\rm kin}/M_*$ as a function of time for models B3H, B3M,
  and B3L (left) and for models B2H, B2M, and B2L (right).
\label{FIG8}}
\end{figure*}

As mentioned above, this instability always sets in for $k \geq 2$
irrespective of magnetic field strength and compactness of the neutron
stars. However, the growth time scale of this instability depends
strongly on the magnetic field strength.  Figure \ref{FIG7} plots
electromagnetic energy as a function of time for several models with
$k=1$ (left) and $k=2$ (right). For models A3H and A2H which are
stable models, it remains approximately constant, whereas for models
with $k=2$, it always decreases during the evolution and the decrease
time scale is shorter for the larger electromagnetic energy.

To determine the growth rate of the dynamically unstable mode, it is
convenient to see $T_{\rm kin}$ because it is initially zero and
increases purely due to excitation of the unstable modes.  Figure
\ref{FIG8} plots $T_{\rm kin}/M_*$ as a function of time for models
B3H, B3M, and B3L (left) and for models B2H, B2M, and B2L
(right). Note that for other unstable models, the behavior of the
curve is qualitatively the same.  This figure shows that after the
onset of the instability, the kinetic energy increases approximately
in an exponential manner with time as $\propto e^{t/\tau}$ where
$\tau$ is a constant.  This indicates that the instability is
dynamical.  This exponential growth holds for all the unstable models
irrespective of the value of $k$ and magnetic field strength.  Thus,
we refer to $\tau$ as the growth time scale.

Table II lists approximate values of $\tau$ for all the unstable
models. It is found that $\tau$ increases systematically with the
decrease of magnetic field strength. In Table I, we also
describe the values of $\bar \tau_A$ calculated using
Eq. (\ref{Alf}). We find that the order of magnitude of $\tau$ agrees with 
$\bar \tau_A$ well, and furthermore, the relation $\tau/\bar
\tau_A \approx 0.3$--0.6 holds irrespective of the value of $k$ and the 
magnetic field strength. This indicates that the dynamical instability
grows in the Alfv\'en time scale.

As we have reported, the instability sets in only for $k\geq 2$ and 
the growth time scale is approximately proportional to the 
Alfv\'en time scale. These imply that this instability is indeed the Tayler 
instability \cite{Tayler,Spruit}. Hereafter, we refer to 
this instability as the Tayler instability. 
 
In this paper, we input a very high magnetic field strength of $\sim
10^{16}$--$10^{17}$ G. Because the scaling holds as shown above, the
present result may be applied for neutron stars of canonical field
strength $\sim 10^{12}$--$10^{13}$ G or magnetar field strength $\sim
10^{14}$--$10^{15}$ G.  As shown in Eq. (\ref{taualf}), the Alfv\'en
time scale is $\sim 10$--100 ms for the magnetar field strength and
$\sim 1$--10 s for the canonical field strength.  Thus, the growth
time scale of the Tayler instability is $\alt 10$ s for the field
strength larger than $\sim 10^{12}$ G. If this instability sets in for 
a neutron star, the electromagentic energy will be redistributed and
transformed into kinetic energy in a short time scale (as short as
or shorter than the rotation period).

\begin{table}[t]
\caption{The growth time, $\tau$, of the Tayler instability for the
  unstable models. The third column denotes the time span of the data
  set which is used for deriving the growth time.  The fourth column
  denotes ratio of $\tau$ to the averaged Alfv\'en time scale derived
  from Eq. (\ref{Alf}).  The error size of the evaluated value of 
$\tau$ is   $\sim 3M_0$.}
\begin{tabular}{cccc} \hline
Model~ & ~~$\tau/M_0$~~ & ~~$t/M_0$~~ & ~~$\tau/\bar \tau_A$~~ \\ \hline
B3H & $\approx 35$  & 200--400 & $\approx 0.48$  \\ 
B3M & $\approx 95$  & 200--600 & $\approx 0.54$ \\
B3L & $\approx 160$ & 200--800 & $\approx 0.61$ \\
B2H & $\approx 40$  & 200-400  & $\approx 0.41$ \\ 
B2M & $ \approx 90$ & 200--800 & $\approx 0.49$ \\
B2L & $\approx 170$ & 200--1000& $\approx 0.57$ \\ 
C3L & $\approx 115$ & 200-700  & $\approx 0.49$ \\ 
C2H & $\approx 32$  & 150--350 & $\approx 0.32$ \\ 
\hline
\end{tabular}
\end{table}

\begin{figure*}[t]
\epsfxsize=2.3in
\leavevmode
\epsffile{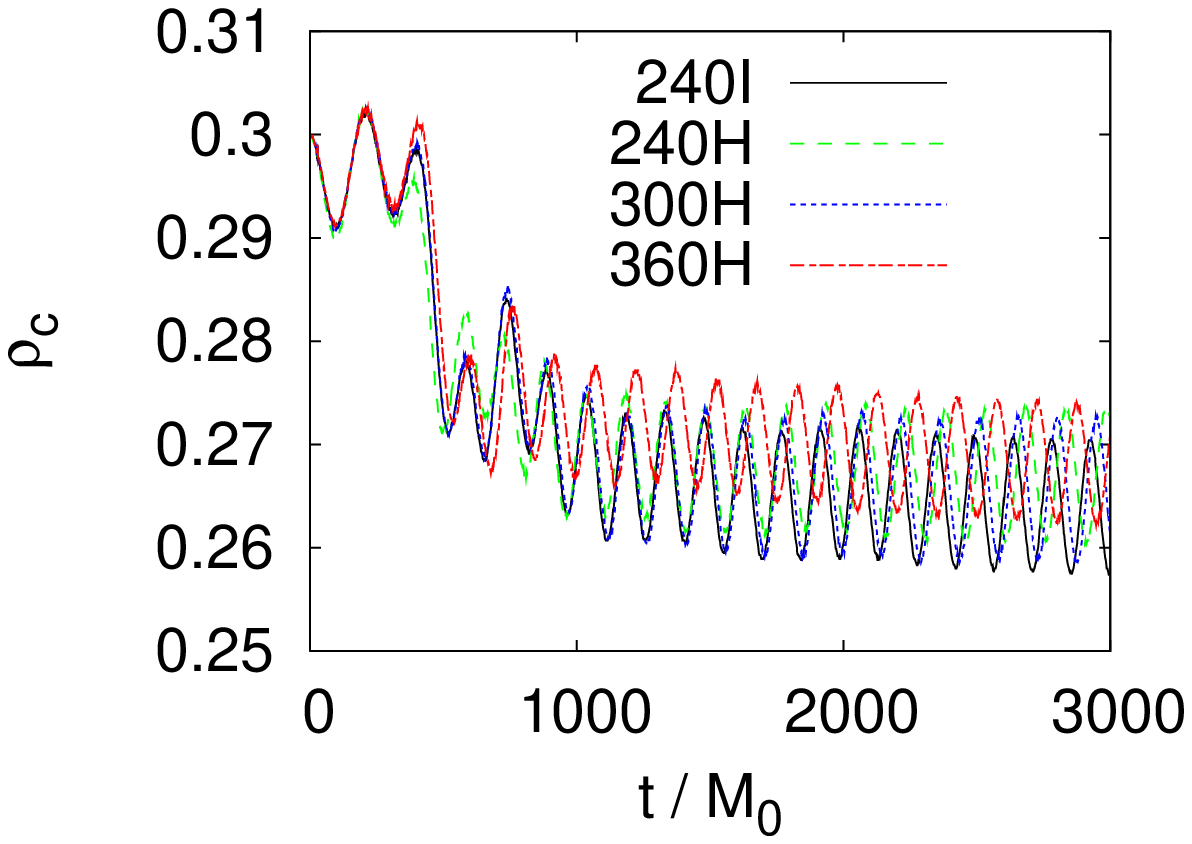}
\epsfxsize=2.3in
\leavevmode
\epsffile{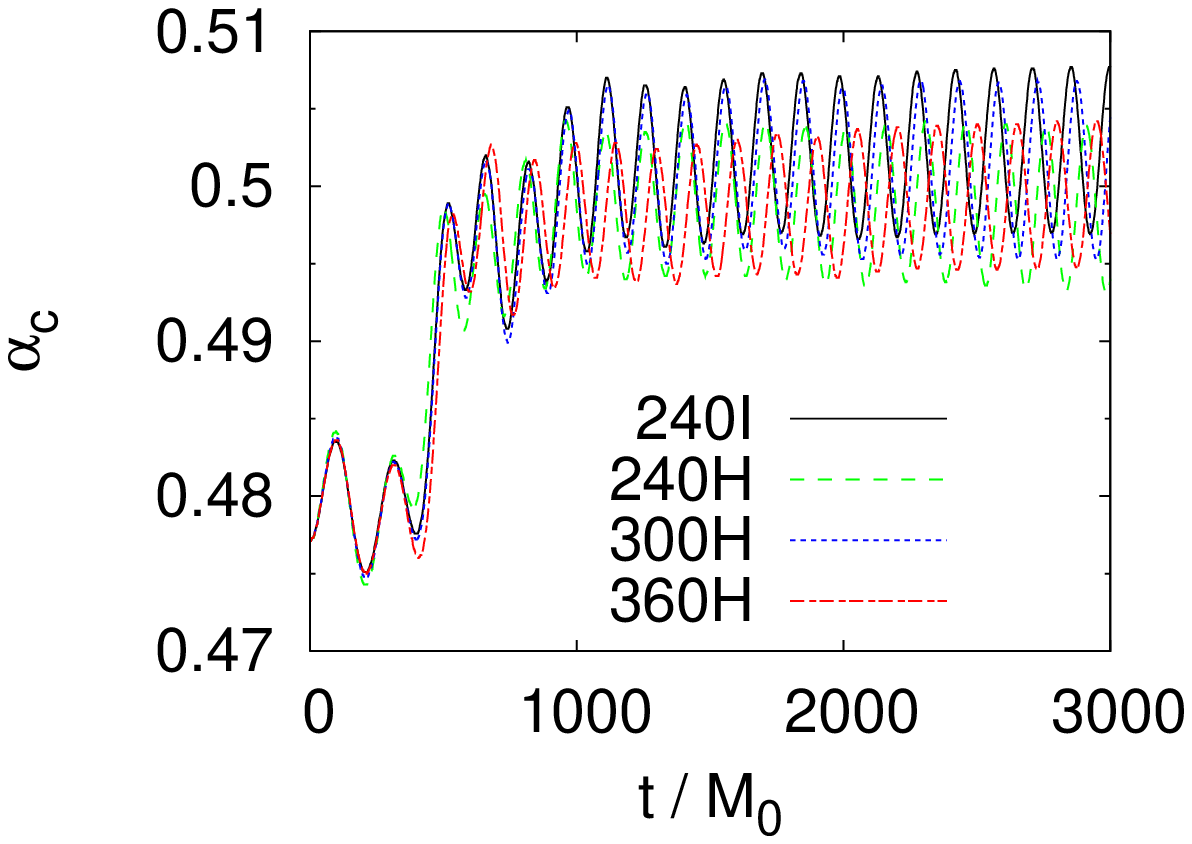}
\epsfxsize=2.3in
\leavevmode
\epsffile{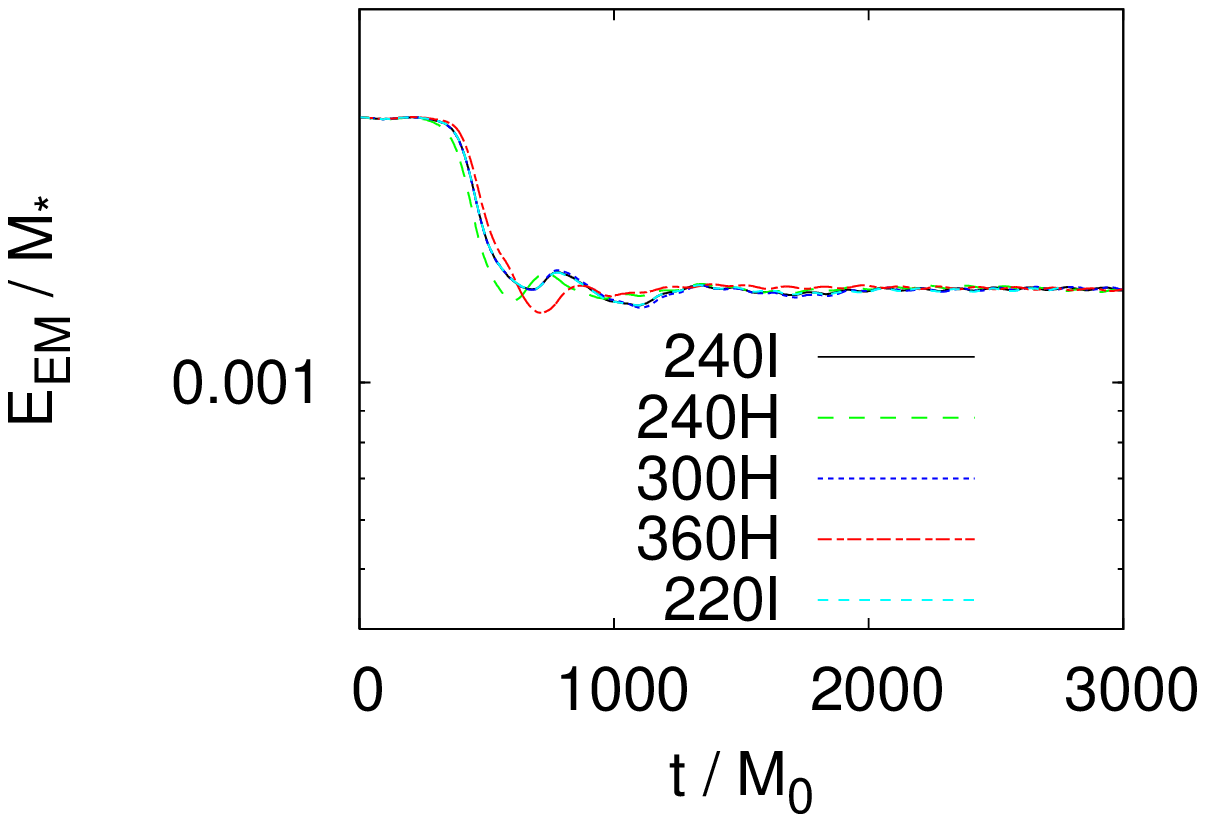}\\
\epsfxsize=2.3in
\leavevmode
\epsffile{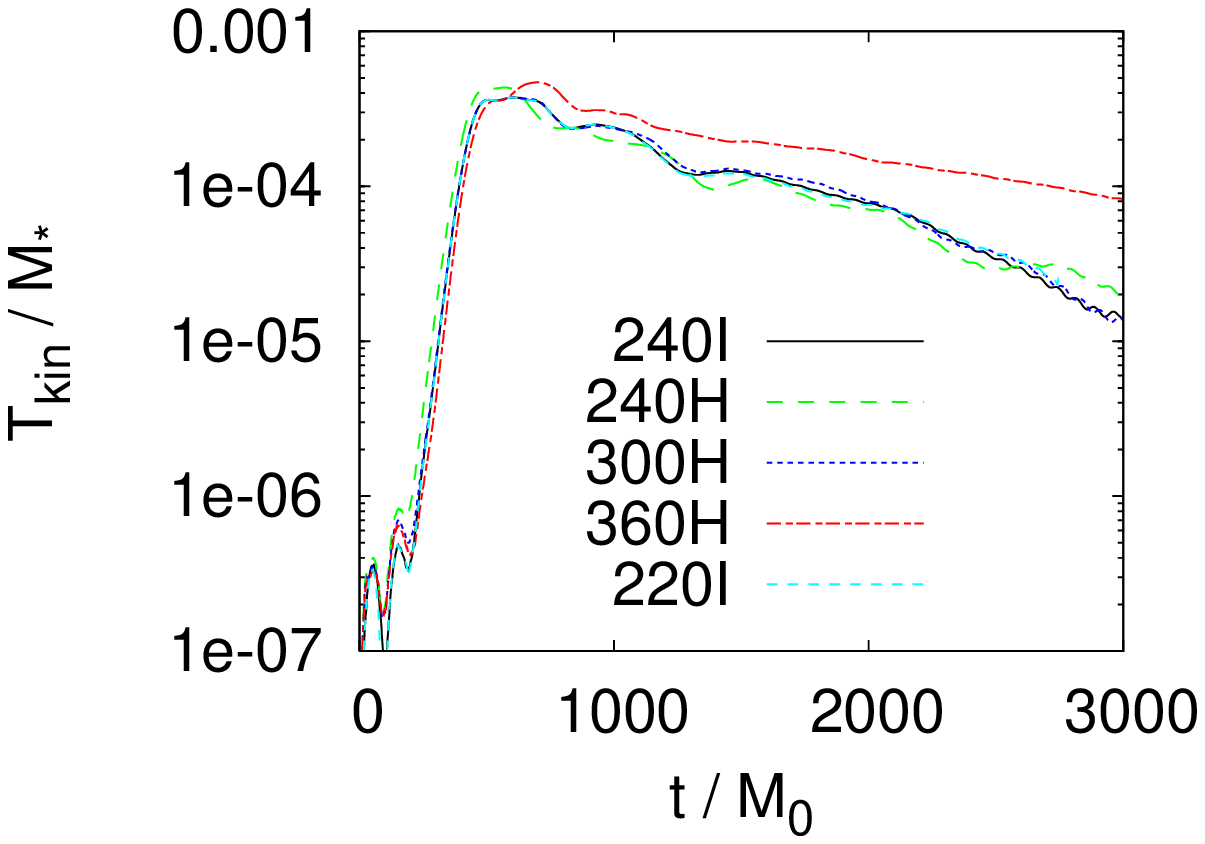}
\epsfxsize=2.3in
\leavevmode
\epsffile{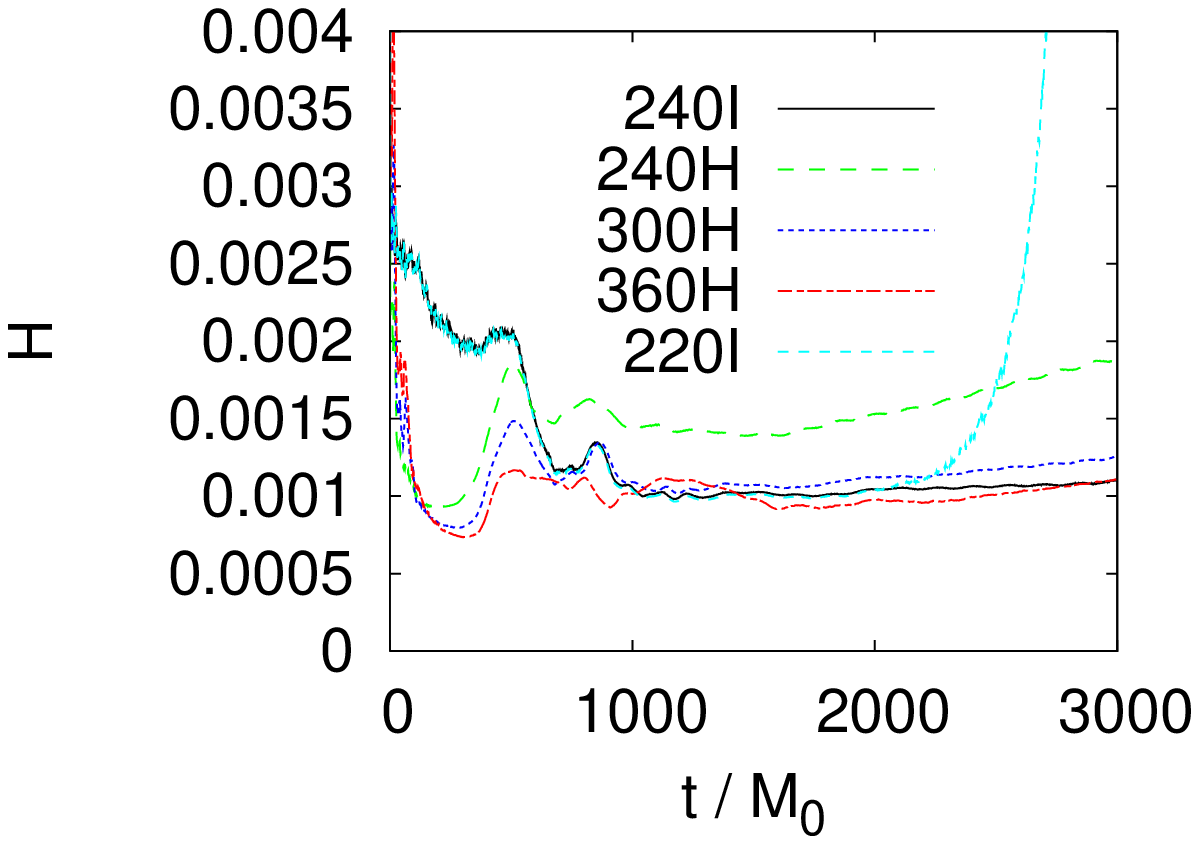}
\epsfxsize=2.3in
\leavevmode
\epsffile{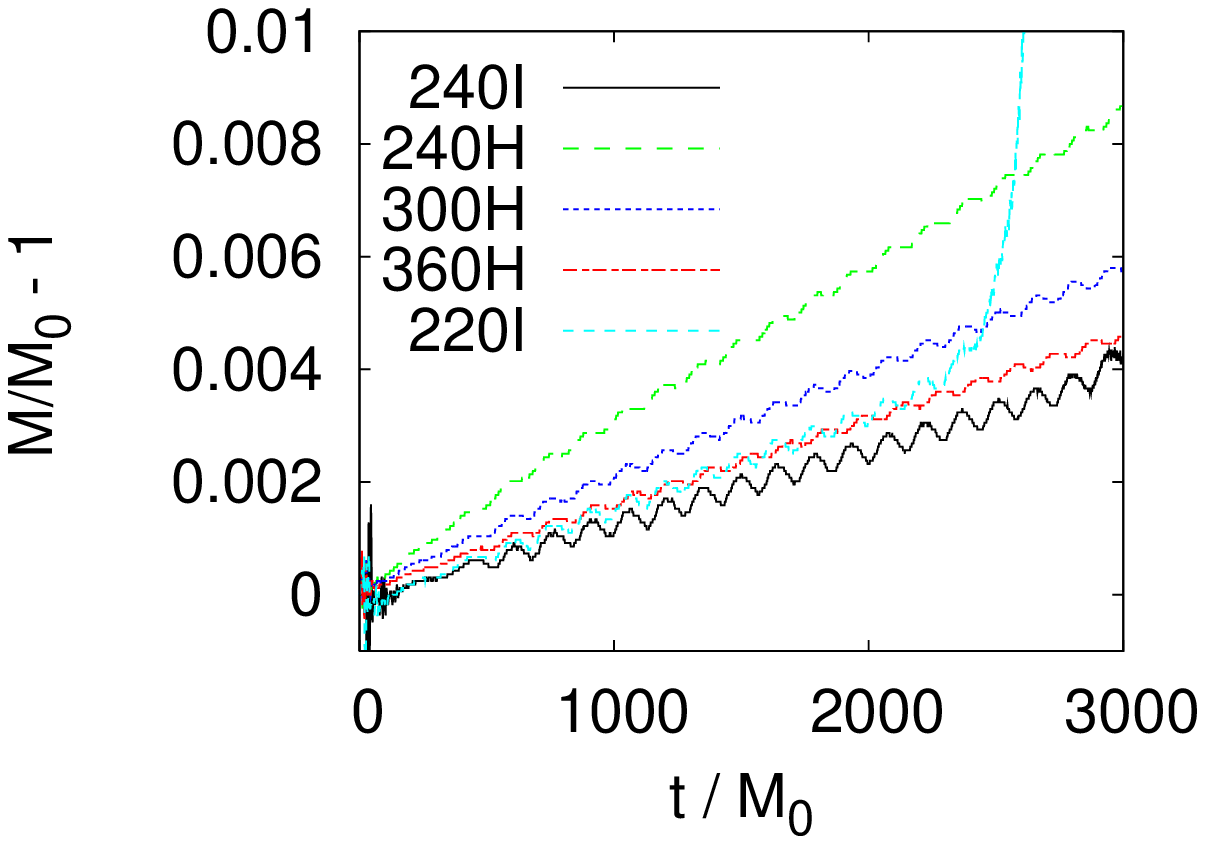}
\vspace{-2mm}
\caption{Evolution of central density, central value of the lapse
  function, electromagnetic energy, kinetic energy, violation of the
  Hamiltonian constraint, and violation of conservation of the ADM
  mass for model B3H with five different grid structures. ``240I'',
  ``220I'', ``240H'', ``300H'', and ``360H'' imply nonuniform grid
  with $N=240$ and $N=220$, uniform grid with $N=240$, 300, and 360,
  respectively.  The units of time is initial ADM mass
  $M_0$. Numerical results for $\alpha_c$ and $\rho_c$ are not shown
  for run 220I because they are approximately the same as those for
  run 240I for $t \alt 2500M_0$. Computation for run 220I crashes 
at $t \sim 2700M_0$. 
\label{FIG9}}
\end{figure*}

For illustrating that the results presented so far depend only weakly
on grid resolution, we show numerical results with different grid
resolutions and grid structures for model B3H. The chosen grid strcuture
is described in Sec. III A. Figure \ref{FIG9} plots the evolution of
the central density, the central value of the lapse function, the
electromagnetic, and kinetic energy, respectively. We also show
violation of the Hamiltonian constraint and conservation of the ADM
mass. Here the definition of the violation of the Hamiltonian
constraint is the same as that shown in Eq. (43) of \cite{S03}; an
averaged Hamiltonian constraint is defined by using the rest-mass
density as a weight.  We note that the ADM mass does not have to be
conserved in axial symmetry, but in the present context with
negligible gravitational radiation, it should be approximately
conserved.

We find that the numerical results are qualitatively the same
irrespective of the grid setting, and also depend quantitatively
weakly on it: At approximately the same time, the neutron star expands
due to the nonlinear growth of the Tayler instability, and then, the
density, the lapse function, and the electromagnetic energy relax to a
stable state (besides oscillation around new equilibrium values). The
minumum electromagnetic energy and the maximum kinetic energy achieved
depends weakly on the grid resolution and differences among five runs
are at most $\sim 20\%$. For $t \alt 1000M_0$, the violation of 
the Hamiltonian constraint is larger for the case that non-uniform 
grid is employed, but this is purely due to the grid structure chosen 
\footnote{Here, let us explain why Hamiltonian constraint violation
  behaves like in Fig. 9(e).  At $t=0$, the degree of violation is much
  larger than 0.3 \%; it is about 1\%.  The reason for this is that
  the initial condition does not satisfy the constraint with a high
  accuracy because the initial condition is prepared by
  linear-interpolating the quantities of the equilibrium solution
  computed in a different grid structure from that in the
  simulation. For $t > 0$, the degree of constraint violation decreases
  by the effect of constraint damping term included in this code. This
  term makes the equation for the conformal factor ($\phi$), which
  primarily determines the Hamiltonian constraint violation, be
  parabolic. Therefore, the equation is not local one and the decrease
  rate of the constraint violation should depend on the grid structure.
} and magnitude of the violation eventually relaxes to be small.
These facts demonstrate that our choice for the grid resolution and 
the grid structure is appropriate for studying this type of
instability.  One caution is that the kinetic energy in the late time
depends strongly on the grid resolution. The likely reason is that
with poorer grid resolutions, numerical viscosity dissipates the
circulation, resulting in the suppress of the convective energy. This
is illustrated by the fact that for the case of ``360H'', the kinetic
energy is largest among five runs. Thus, in reality, the convective
motion may be approximately constant for a time much longer than the
Alfv\'en time scale.  However, accurately following the convective 
motion for a long time is not main subject of this paper, and hence, 
we do not touch this problem in detail.

\subsubsection{Rotating case}

\begin{figure*}[t]
\epsfxsize=3.in 
\leavevmode 
\epsffile{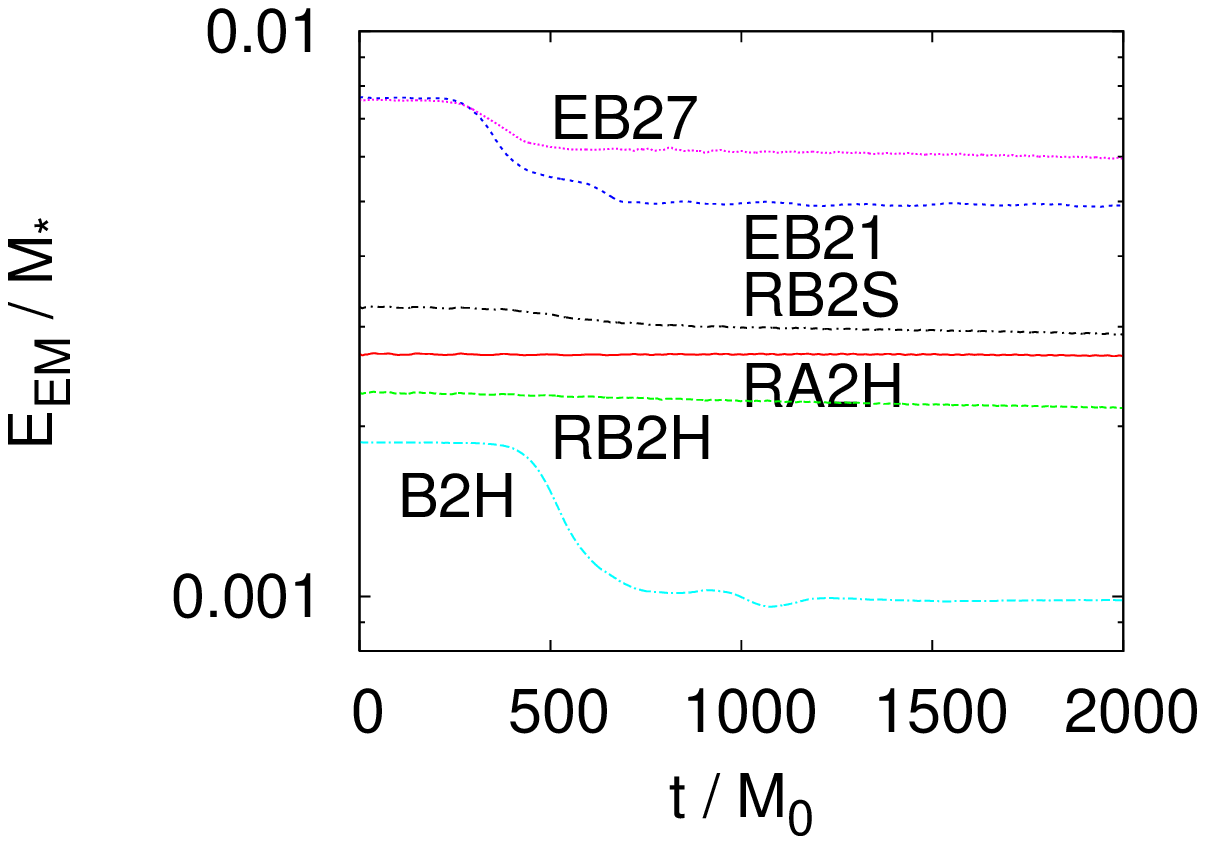}
\epsfxsize=3.in 
\leavevmode 
~~~~\epsffile{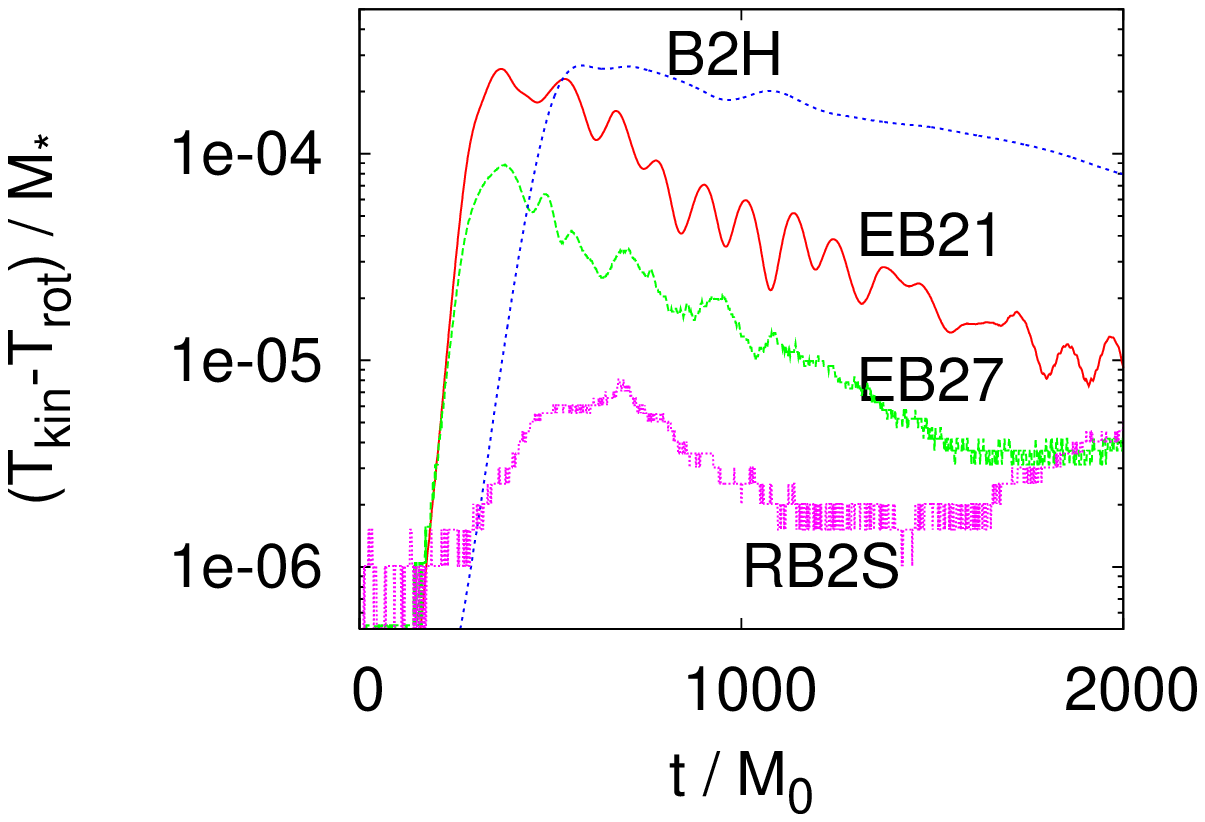}
\caption{$E_{\rm EM}/M_*$ and $(T_{\rm kin}-T_{\rm rot})/M_*$ as 
  functions of time for models RA2H, RB2H, RB2S, EB21, and EB27. For
  comparison, results for model B2H are plotted together. 
For models RA2H and RB2H, $(T_{\rm kin}-T_{\rm rot})/M_*$ remains 
to be smaller than $5 \times 10^{-7}$. 
\label{FIG10}}
\end{figure*}

\begin{figure}[th]
\epsfxsize=3.4in 
\leavevmode 
\epsffile{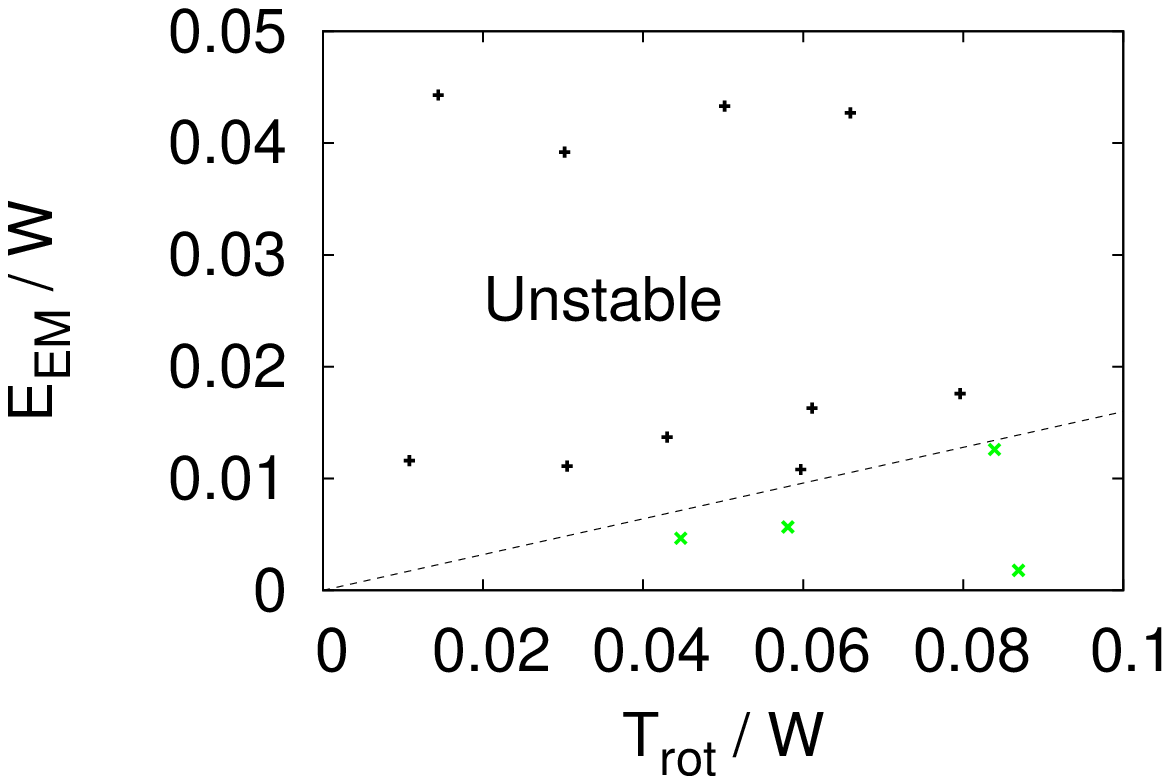}
\vspace{-5mm}
\caption{Results of stability properties for rotating models with
  $k=2$. All the stars above the dashed line are unstable, and the
  instability grows in the time scale of $\alt \bar \tau_A$. For the
  stars below the dashed line, instability does not grow in the
  Alfv\'en time scale, although we cannot exclude the possibility that the
  stars are unstable and the unstable modes grow for a time scale $\gg
  \bar \tau_A$.
\label{FIG11}}
\end{figure}

Numerical simulations for rigidly rotating neutron stars were
performed for all the models listed in Table I. We find that the
stability criteria for the rotating stars are different from those for
nonrotating stars. As in the nonrotating case, stars with $k=1$ are
always stable against axisymmetric perturbation irrespective of
magnetic field strength. Also, for many of stars with $k=2$, the
Tayler instability occurs and grows approximately in the Alfv\'en time
scale.  In contrast to the nonrotating case, however, stars with $k
\geq 2$ may be stable for the rapidly rotating case; at least, for the
first $\sim 10$ Alfv\'en time scale, we do not find evidence for
occurring the Tayler instability.  Figure \ref{FIG10} plots evolution
of electromagnetic and convective kinetic energy (defined by $T_{\rm
  kin}-T_{\rm rot}$) for models RA2H, RB2H, RB2S, EB21, and EB27.
This shows that for the rapidly rotating models, RA2H and RB2H, the
electromagnetic energy remains approximately constant and indicates
that they are stable, at least, in the time scale of $t = 2000M_0$,
more than $10\bar \tau_A$ 
\footnote{We cannot exclude a possibility that the rapidly rotating
  stars are eventually unstable after a longterm evolution. As shown
  in \cite{Acheson,Tayler2}, the growth rate is proportional to
  $v_A^2/R^2\Omega$ for the rapidly rotating case, and hence, it might
  be necessary to perform simulation for a much longer time scale. }.
It is worth noting that for model RB2H, $k=2$ and electromagnetic
energy is strong as $E_{\rm EM}/W \sim 0.01$. Nevertheless, it is a
stable model: This illustrates that rapid rotation suppresses the
Tayler instability.  By contrast, models RB2S and EB27, which have
even larger electromagnetic energy $E_{\rm EM}/W \sim 0.02$--0.04, are
unstable even for a large value of $T_{\rm rot}/W \sim 0.07$.

These results indicate that the stability is determined by the ratio
$E_{\rm EM}/T_{\rm rot}$; i.e., the Tayler instability grows in the
Alfv\'en time scale only when $E_{\rm EM}/T_{\rm rot}$ is larger than
a critical value $\sim 0.2$. To clarify this fact, we generate
Fig. \ref{FIG11} which summarizes the stability properties for
rotating models with $k=2$. This figure indeed suggests that the
Tayler instability grows only for $E_{\rm EM}/T_{\rm rot} \agt 0.2$
(note that the dashed line denotes $E_{\rm EM}/T_{\rm
  rot}=0.16$). This result is in qualitative agreement with the
Newtonian analysis of Appendix A.

As seen in Fig. \ref{FIG10}, the fraction of the decrease in the
electromagnetic energy during evolution is smaller for larger value of
$T_{\rm rot}/W$. Also, shown is that for larger ratio of $T_{\rm
  rot}/E_{\rm EM}$, the maximum value of convective kinetic
energy, $T_{\rm kin}-T_{\rm rot}$, is smaller and the convective motion
damps more quickly for the rotating models. In particular, 
for the rapidly rotating case, the convective kinetic energy 
is smaller than the electromagnetic energy by two or three 
orders of magnitude. As mentioned in the
previous section, the damping is partly due to the numerical
dissipation.  However, significant difference in the damping rates
between nonrotating and rotating models suggests that the damping is
primarily due to a rotational effect. All these facts indicate that
rotation plays a role for stabilizing the axisymmetric unstable mode. 

As discussed in \cite{Tayler2,Spruit}, the Tayler instability induces
a motion perpendicular to the rotation axis, and stabilization by
rapid rotation is due to the presence of the Coriolis force.  The
Coriolis force pushes a fluid element, which deviates from its
equilibrium location to the positive cylindrical-radial direction, to
the counter-rotation direction.  As a result, the centrifugal force of
the fluid element is weaken, and it is enforced to
return toward its equilibrium position.  Thus, it is natural that the
Coriolis force suppresses the onset of the Tayler instability as well
as convective motion in the meridian plane.

\section{Summary}

In this paper, we have reported stability of neutron stars with
purely toroidal magnetic fields. The stability is determined by
performing GRMHD simulation. For the simulation, we prepare a variety
of equilibrium neutron stars changing their compactness, strength and
profile of toroidal magnetic fields, and rotational kinetic
energy. The following is the summary of the numerical results.

\noindent
(i) Magnetized stars with $k=1$ are stable against axisymmetric perturbation 
irrespective of the magnetic field strength and the rotational 
kinetic energy. 

\noindent
(ii) For the nonrotating case with $k=2$, magnetized stars are dynamically
unstable irrespective of the magnetic field strength, and the magnetic
field is redistributed approximately in the Alfv\'en time
scale. The resulting profile of the magnetic fields is similar to
that for $k=1$, indicating that stars with $k=1$ are attractors if
only the axisymmetric perturbation is taken into account.

\noindent
(iii) During the growth of the dynamically unstable modes,
electromagnetic energy is transported to the kinetic energy via the
Tayler instability, and as a result, a convective motion is
excited. The magnitude of the convective kinetic energy becomes
approximately as large as the electromagnetic energy at the time when
the nonlinear growth saturates, for not-rapidly rotating neutron stars. 

\noindent
(iv) For the case that a neutron star is rapidly rotating, the Tayler
instability is suppressed and the star with $k=2$ may be dynamically
stable against axisymmetric perturbation at least in about ten
Alfv\'en time scale. The numerical results suggest that if the
rotational kinetic energy is more than $\sim 6$ times larger than the
electromagnetic energy, the star is stabilized.

\noindent
(v) Even for the unstable rotating models, convective motion is not
induced as remarkablely as in the nonrotating case. For the rapidly
rotating case in which the rotational kinetic energy is larger than
the electromagnetic energy by a factor of $\agt 2$, the maximum
convective energy is smaller than the electromagnetic energy by two or
three orders of magnitude. This indicates that rotation in general
plays a role in stabilizing the axisymmetric Tayler instability.

As mentioned in Sec. I, protoneutron stars are likely to have strong
toroidal magnetic fields if the progenitor of supernova gravitational
collapse is rotating. Indeed, a number of recent MHD simulations of
supernova collapse of magnetized rotating stars have shown that the
toroidal magnetic field is amplified by many order of magnitude during
collapse and during subsequent relaxation stage of the formed
protoneutron star
\cite{YS,SKY,TKNS,KOTAKE,OAM,OADM,ABM,MBA,SLSS,DBLO,DFD}. The key
mechanism in this amplification is transport of the rotational kinetic
energy to the electromagnetic energy via winding induced by
differential rotation. As indicated in \cite{SLSS}, the
electromagnetic energy could be comparable to the rotational kinetic
energy at the end of the amplification. Our present numerical
experiment suggests that when the condition $E_{\rm EM}/T_{\rm rot}
\agt 0.2$ is achieved, the protoneutron star may be subject to the
Tayler instability.

Many of the MHD studies for supernova collapse focus on the
amplification of magnetic pressure associated with the amplification
of the toroidal magnetic field strength, which increases the pressure
behind shock waves and can help supernova explosion or drive a strong
outflow along the rotational axis.  Our present study suggests that
the strong toroidal magnetic field may play a role not only in increasing
the pressure for pushing shock waves but also in exciting a convective
motion through the Tayler instability.  After the onset of this
instability, a large fraction of electromagnetic energy may be
transported into the convective kinetic energy. Then the convection may
help to carry a hot material near the surface of a protoneutron star
toward the gain region and push stalled shock waves outward
\cite{Janka}.  As Eq. (\ref{conv}) indicates, the kinetic energy of
the convective motion may increase to $\sim 10^{50}~{\rm ergs}$ if the
toroidal field strength becomes $10^{16}$ G and the electromagnetic
energy becomes as large as the rotational kinetic energy. This value of the
kinetic energy amounts to $\sim 10\%$ of the energy required to drive
supernova explosion. As we show in Fig. \ref{FIG5}, vorticity is
generated associated with the convective motion. If this vorticity is
dissipated by viscosity, a large thermal energy is also generated.
Such thermal energy may also contribute to pushing stalled shock waves
(see similar discussion in \cite{TQB}). All these possibilities
suggest that the Tayler instability should be taken into account
in magnetorotational explosion scenarios for supernova explosion. 

In reality, electromagnetic energy in protoneutron star at birth is
likely to be much smaller than rotational kinetic energy. Thus, at its
birth, the protoneutron star will be stable against the Tayler
instability. Subsequently, the toroidal magnetic fields are
amplified by winding caused by differential rotation, and as a result,
the electromagnetic energy will reach a magnitude comparable to the
rotational kinetic energy. Then, the protoneutron star could be
unstable against the Tayler instability. This suggests that this
instability may play a role for stopping the growth of the toroidal
field strength.  At the end of the toroidal field amplification, the
resulting electromagnetic energy is likely to be at most comparable to
the rotational kinetic energy, and hence, the instability may not be
as strong as that in the nonrotating stars, as illustrated in Sec. III
B. Also, the convective energy driven by the instability is likely to
depend strongly on the amplification process of the toroidal magnetic
fields via winding. For example, if the degree of differential
rotation is largest near the rotation axis, the instability will not
be strong because the configuration is likely to be similar to that of
$k=1$.  On the other hand, the degree of differential rotation is
largest at an interior of a protoneutron star far from the rotation
axis, the Taylor instability will occur. The efficiency of the winding
depends strongly on the magnetic field profile and the rotational profile
of progenitor.  To understand the role of the Tayler instability,
well-resolved, long-term MHD simulations of stellar core collapse have
to be systematically performed for a number of initial conditions.

Neutron stars, observed as ordinary pulsars, typically have 
rotation period 0.1--1 s and magnetic field strength of  
$10^{11}$--$10^{13}$ G. This class of neutron stars are not 
subject to the Tayler instability. This is because the 
ratio of electromagnetic energy to rotational 
kinetic energy is much smaller than unity as
\beqn
{E_{\rm EM} \over T_{\rm rot}}
&\sim &{B^2 R^3/3 \over I \Omega^2/2} \nonumber \\
&\approx & 7 \times 10^{-3}
\biggl({B \over 10^{13}~{\rm G}}\biggr)^2
\biggl({R \over 10~{\rm km}}\biggr)^{3} \nonumber \\
&&~~~~~~ \times \biggl({I \over 10^{45}~{\rm g~cm^2}}\biggr)^{-1}
\biggl({\Omega \over 10~{\rm rad/s}}\biggr)^{-2},
\eeqn
where $I$ is moment of inertia. 

By contrast, magnetars of rotation period 5--12 s and of magnetic
field strength $10^{14}$--$10^{15}$ G \cite{WD} are subject to the
Tayler instability. Present numerical results imply that stable 
magnetars should have a magnetic field profile which is at least
stable against axisymmetric perturbation.

In this paper, we focus only on the stability against axisymmetric
perturbations. This work should be regarded as the first step toward
deeper understanding of the Tayler instability in magnetized neutron
stars.  As shown in \cite{Tayler,Acheson,Tayler2,Spruit}, neutron
stars with toroidal magnetic fields are also unstable against
nonaxisymmetric perturbation. Nonaxisymmetric Tayler instability may
grow in a different manner from axisymmetric one, and hence, the
dynamical evolution process of neutron stars during the growth of the
unstable modes as well as the final fate could also be different.
Furthermore, this instability will occur even for $k=1$. In the
axisymmetric simulation, only the stars with $k \geq 2$ are unstable
and magnetic field profile of such unstable star eventually relaxes to
a profile similar to that with $k=1$. However, in three dimensions,
such star will be still unstable. This suggests that stars may never
reach a stationary state.  To answer this question, GRMHD simulation
in full three dimensions is required.  We plan to perform three
dimensional simulation in the next step.

\acknowledgments

We thank T. Suzuki for helpful discussion.  Numerical computations
were performed in part on the NEC-SX8 at Yukawa Institute of
Theoretical Physics of Kyoto University. This work was in part
supported by Monbukagakusho Grant (Nos. 19540263 and 19540309).
 
\appendix

\section{Perturbative analysis}
In this section, we present a result of perturbative analysis on
criteria for the onset of axisymmetric instabilities of neutron stars
with toroidal magnetic fields. For the analysis, we follow
Ref. \cite{Acheson}.  Newtonian gravity is assumed for the sake of
clarity and simplicity, and thus, our purpose is to derive an
approximate criterion for the instabilities.
\footnote{In this Appendix, we recover the gravitational constant $G$. 
Also, $c$ does not denote the speed of light.}

Basic equations describing ideal MHD are given by 
\begin{eqnarray}
&&\partial_t\rho+\nabla_i(\rho v^i)=0\,,
\label{eq:mass_conservation}\\
&&\rho\,(\partial_t v_i+v^j\nabla_j v_i)=-\nabla_i\left(P+{1\over 
  8\pi}\, B^j B_j\right) \nonumber \\ 
&&~~~~~~~~~~~~~~~~~~~~~~~~+{1\over 4\pi}\,B^j\nabla_j B_i-\rho g_i^*\,,
\label{eq:Euler}\\
&&\partial_t B^i=\nabla_j(v^i B^j-v^j B^i)\,, 
\label{eq:induction}\\
&&\nabla_i B^i=0\,,
\end{eqnarray}
where $\rho$ denotes the rest-mass density, $v^i$ the fluid velocity,
$P$ the pressure, $B^i$ the magnetic field, $g^*_i$ the gravitational
acceleration, and $\nabla_i$ the covariant derivative with respect to
$x^i$.

We derive linear perturbation equations for rigidly rotating stars
with purely toroidal magnetic fields in equilibrium. The 
velocity and the magnetic field for the equilibrium stars 
are written as
\begin{eqnarray}
v^i&=&(0,\Omega,0)\,,\\
B^i&=&\left(0,\varpi^{-1}B(\varpi,z),0\right)\,,
\end{eqnarray}
where $\Omega$ and $B(\varpi,z)$ are the angular velocity and the
magnetic field strength, respectively.  Here, we used the
cylindrical polar coordinates $(\varpi,\varphi,z)$. 

In order for the stability analysis to be tractable, we only consider
axisymmetric perturbations of very short wavelength both in the
$\varpi$ and $z$ directions. Here, the short wavelength implies that
the wavelength, $\lambda$, of an oscillation mode is smaller than
$\delta v \tau$ where $\delta v$ is a typical magnitude of the
perturbed velocity field and $\tau$ is a typical change time scale of
stellar structure.  In other words, we perform the local analysis.  We
also employ the Cowling approximation, in which perturbations of the
gravity are omitted.

In the short-wavelength approximation, linear perturbation equation for 
the mass conservation equation (\ref{eq:mass_conservation}) is 
\begin{eqnarray}
\partial_\varpi \delta v^\varpi+\partial_z\delta v^z=0\,, 
\label{eq:p_mass_conservation}
\end{eqnarray}
where $\delta Q$ denotes the Euler perturbation of the physical
quantity $Q$ and we assume that $|\pa_\varpi \delta v^{\varpi}| \gg
|\delta v^{\varpi}/\varpi|$ because of the short-wavelength
approximation.  Equation (\ref{eq:p_mass_conservation}) implies that
the effect of density perturbation does not play a role. This is
also because of the short-wavelength approximation imposed. Thus,
sound waves or $p$-modes are filtered out in this analysis. 

In the short-wavelength approximation, 
the $\varpi$ and $z$ components of Eq. (\ref{eq:Euler}) become the
following same equation
\begin{eqnarray}
\delta P+{1\over 4\pi}\, \varpi B\,\delta B^\varphi=0\,. 
\label{eq:p_euler1}
\end{eqnarray}
The other pieces of independent information extracted from Eq. 
(\ref{eq:Euler}) are given by
\begin{eqnarray}
&&\partial_t\delta v_\varphi+2\varpi\Omega\,\delta v^\varpi
={1\over 4\pi}\,\delta B^j\partial_j(\varpi B)\,, 
\label{eq:p_euler2}\\
&&\rho [\partial_t(\partial_z\delta v_\varpi-\partial_\varpi\delta v_z)-
2\varpi\Omega\,\partial_z\delta v^\varphi] \nonumber \\
&&~~~~~={1\over 4\pi}\,(-2 B\,\partial_z\delta B^\varphi)
-g_\varpi\partial_z\delta\rho+g_z\partial_\varpi\delta\rho\,,~~~
\label{eq:p_euler3}
\end{eqnarray}
where $g_i$ is the apparent gravity, defined by 
\begin{eqnarray}
g_i\equiv g_i^*+u^j\nabla_j u_i=(g^*_\varpi-\varpi\Omega^2,0,g^*_z)\,. 
\end{eqnarray}

The induction equation (\ref{eq:induction}) gives 
\begin{eqnarray}
&&\partial_t\delta B^\varpi=0\,,\quad \partial_t\delta B^z=0\,,
\label{eq:p_induction1}\\
&&\partial_t\delta B^\varphi={B\over\varpi\rho} \,\partial_t\delta\rho-
\rho\,\delta v^j\partial_j\left({B\over\varpi\rho}\right)\,. 
\label{eq:p_induction2}
\end{eqnarray}
In this study, we focus on adiabatic oscillations. Then, the
relationship between $\delta P$ and $\delta \rho$ is 
\begin{eqnarray}
\delta P+\xi^i\partial_i P={P\Gamma\over
  \rho}\,(\delta\rho+\xi^i\partial_i \rho)\,,
\label{eq:p_adiabatic0}
\end{eqnarray}
where $\Gamma$ denotes the adiabatic index, defined by 
\begin{eqnarray}
\Gamma\equiv \left({\ln P \over \ln\rho}\right)_{\rm ad}\,,
\end{eqnarray}
and $\xi^i$ the Lagrangian displacement, which obeys 
\begin{eqnarray}
\delta v^i=\partial_t \xi^i+v^j\partial_j \xi^i-\xi^j\partial_j v^i\,.
\label{A17}
\end{eqnarray}
For the axisymmetric perturbation with $v^j=\Omega \delta_{\varphi}^{~j}$, 
Eq. (\ref{A17}) reduces to 
\begin{eqnarray}
\delta v^i=\partial_t \xi^i\,.
\end{eqnarray}
Then, Eq. (\ref{eq:p_adiabatic0}) becomes 
%
\begin{eqnarray}
{1\over v_s^2\rho}\,\partial_t \delta P=
{1\over \rho}\,\partial_t \delta\rho+\delta v^iA_i\,,
\label{eq:p_adiabatic}
\end{eqnarray}
where $v_s$ is the adiabatic sound speed, defined by 
\begin{eqnarray}
v_s\equiv\left({P\Gamma\over\rho}\right)^{1\over2}\,,
\end{eqnarray}
and $A_i$ the Schwarzschild discriminant, 
\begin{eqnarray}
A_i\equiv\partial_i\ln\rho-{1\over\Gamma}\,\partial_i\ln P\,.
\end{eqnarray}

As shown in Eq. (\ref{eq:p_induction1}), we have $\delta
B^\varpi=0=\delta B^z$ because we are not interested in
time-independent perturbations.  Thus, Eqs. 
(\ref{eq:p_mass_conservation})--(\ref{eq:p_euler3}),
(\ref{eq:p_induction2}), and (\ref{eq:p_adiabatic}) are six
independent equations for the six independent variables $\delta v^i$,
$\delta B^\varphi$, $\delta\rho$, and $\delta P$. In the local 
analysis, axisymmetric perturbations can be written as
\begin{eqnarray}
\delta Q(t, \varpi, z)=Q_0 \exp\{i(-\sigma t+l\varpi+nz)\}\,,
\label{eq:Fourier_transformation}
\end{eqnarray}
where $Q_0$ is a constant, $\sigma$ the oscillation frequency, 
and $(l, n)$ the meridional wavenumber vector. 

Substituting Eq. (\ref{eq:Fourier_transformation}) into 
the perturbation equations, we obtain 
the following dispersion relation for $\sigma$: 
\begin{eqnarray}
&&\left(1+{v_A^2\over v_s^2}\right){s^2\over n^2}\,\sigma^2=
  4\Omega^2\left(1+{v_A^2\over v_s^2}\right) -\left(\hat
  g+{2v_A^2\over \varpi}\right)A_h \nonumber \\ &&
  ~~~~~~~~~~~~~~~~~~~~~+\left(\hat g-{2v_s^2\over
    \varpi}\right){v_A^2\over v_s^2} {\partial\over \partial
    q}\ln\left({B\over\varpi\rho}\right)\,.
\label{eq:dispersion_relation}
\end{eqnarray}
Here, $v_A$ denotes the Alfv{\'e}n speed 
\begin{eqnarray}
v_A\equiv\left({B^2\over 4\pi\rho}\right)^{1\over 2}\,, 
\end{eqnarray}
$s$ the total meridional wavenumber 
\begin{eqnarray}
s\equiv(l^2+n^2)^{1\over 2}\,, 
\end{eqnarray}
$\hat g$ the apparent gravity along the constant phase or crests
\begin{eqnarray}
\hat g\equiv g_\varpi-{l\over n}\,g_z\,, 
\label{def:G}
\end{eqnarray}
$A_h$ the Schwarzschild discriminant along the constant phase or crests
\begin{eqnarray}
A_h\equiv A_\varpi-{l\over n}\,A_z\,, 
\end{eqnarray}
and $\partial /\partial q$ the derivative along the constant phase or crests 
\begin{eqnarray}
{\partial \over \partial q}\equiv{\partial\over \partial
  \varpi}-{l\over n}\,{\partial\over \partial z}\,.
\label{def:dh}
\end{eqnarray}
The first, the second, and the third terms in the right-hand side of
Eq. (\ref{eq:dispersion_relation}) are related to effects of the
rigid rotation, the stratification (the buoyancy), and the magnetic
buoyancy, respectively. Here, it should be emphasized that the effect
of rigid rotation operates as a stabilizing agent because the first
term in the right-hand side of Eq. (\ref{eq:dispersion_relation})
is always positive. 

The stabilities are determined by the dispersion relation
(\ref{eq:dispersion_relation}). Specifically, the sign of its
right-hand side determines the local stabilities; the stars are
locally stable (unstable) if $\sigma^2>0$ ($\sigma^2<0$).  From
Eqs. (\ref{def:G})--(\ref{def:dh}), we find that the right-hand 
side of Eq. (\ref{eq:dispersion_relation}) is a quadratic in
$l/n$, given by
\begin{eqnarray}
\left(1+{v_A^2\over v_s^2}\right){s^2\over n^2}\,\sigma^2= a \left({l\over n}\right)^2
+b \left({l\over n}\right)+c \,, 
\label{eq:dispersion_relation2}
\end{eqnarray}
where 
\begin{eqnarray}
a&\equiv &g_z \left\{-A_z+{v_A^2\over v_s^2}
{\partial\over \partial z}\ln\left({B\over\varpi\rho}\right)\right\}\,, \\
b&\equiv&\left(g_\varpi+{2v_A^2\over \varpi}\right)A_z+g_zA_\varpi \nonumber \\
&&~~~~~~~ -\left(g_\varpi-{2v_s^2\over \varpi}\right){v_A^2\over v_s^2}
{\partial\over \partial z}\ln\left({B\over\varpi\rho}\right) \nonumber \\
&&~~~~~~~ -g_z{v_A^2\over v_s^2}
{\partial\over \partial \varpi}\ln\left({B\over\varpi\rho}\right)\,,\\
c&\equiv&4\Omega^2\left(1+{v_A^2\over v_s^2}\right)-
\left(g_\varpi+{2v_A^2\over \varpi}\right)A_\varpi \nonumber \\
&&~~~~~~~~ +\left(g_\varpi-{2v_s^2\over \varpi}\right){v_A^2\over v_s^2}
{\partial\over \partial\varpi}\ln\left({B\over\varpi\rho}\right)\,.
\end{eqnarray}
We therefore see that the stability condition $\sigma^2>0$ for any value of $l/n$ is 
equivalent to the condition that  the three inequalities $a>0$, $c>0$, and 
$b^2-4ac<0$ are simultaneously satisfied. Contrapositively, 
it is found that the star is unstable if any of the following three conditions is satisfied:
\begin{eqnarray}
&&4\Omega^2\left(1+{v_A^2\over v_s^2}\right)-
\left(g_\varpi+{2v_A^2\over \varpi}\right)A_\varpi \nonumber \\
&& ~~~~~~~ +\left(g_\varpi-{2v_s^2\over \varpi}\right){v_A^2\over v_s^2}
{\partial\over \partial\varpi}\ln\left({B\over\varpi\rho}\right)<0\,, 
\label{ineq1}\\
&&g_z \left\{-A_z+{v_A^2\over v_s^2}
{\partial\over \partial z}\ln\left({B\over\varpi\rho}\right)\right\}<0\,, 
\label{ineq2}\\
&&g_z\biggl[{v_A^2}\left\{
A_z{\partial\over\partial \varpi}\ln\left({B\over\varpi\rho}\right)
-A_\varpi{\partial\over\partial z}\ln\left({B\over\varpi\rho}\right)
\right\} \nonumber \\
&&~~~~~+{2\varpi\Omega^2}\left\{{v_A^2\over v_s^2}{\partial\over\partial z}
\ln\left({B\over\varpi\rho}\right)-A_z\right\}
\biggr] <0 \,. \label{ineq3}
\end{eqnarray}
Here, the final equation (\ref{ineq3}) is equivalent to the condition
$b^2-4ac>0$.

Neutron stars are likely to be stably stratified because of their
strong composition gradient~\cite{Reisenegger:1992}. As a result, the
buoyancy inside the neutron star exerts as a stabilizing force as
shown in Eqs. (\ref{ineq1}) and (\ref{ineq2}).  Equation (\ref{ineq1})
also shows that the criterion of the magnetic instability depends on
whether the region considered is located inside or outside the
critical surface whose cylindrical radius is defined by
\begin{eqnarray}
 \varpi_c\equiv{2v_s^2\over g_\varpi}\,. 
\end{eqnarray}
Inside (Outside) the critical surface, if ${\partial\over
  \partial\varpi}\ln\left({B/\varpi\rho}\right)>0$ (${\partial\over
  \partial\varpi}\ln\left({B/\varpi\rho}\right)<0$), the third term of
Eq. (\ref{ineq1}) becomes negative and the instability is promoted.

Finally, we examine the magnetic stability of a particular model, a
slowly rotating star containing toroidal magnetic fields. We assume
that matter distribution of the star is spherical (namely the magnetic and
centrifugal forces are not strong enough to modify this spherical
shape).  As an example, an $n=1$ polytropic sphere is considered
because the analytic solution is available. Its matter distribution is
given by
\begin{eqnarray}
\rho&=&\rho_0\left({\sin(r/r_0)\over (r/r_0)}\right)\,, \\
P&=&P_0\left({\sin(r/r_0)\over (r/r_0)}\right)^2\,,
\end{eqnarray}
where $\rho_0$ and $P_0$ are the central values of the density and 
the pressure, respectively, and $r_0$ is stellar radius 
$$
r_0\equiv\left({2 P_0\over 4\pi G\rho_0^2}\right)^{1\over 2}\,.
$$ 
For the magnetic field distribution, 
we take a simple form, as in \cite{KY}, as 
\begin{eqnarray}
B=b_0 (\rho/\rho_0)^k (r\sin\theta/r_0)^{2k-1}\,, 
\label{def:B}
\end{eqnarray}
where $b_0$ and $k$ are constants. Regularity of the magnetic fields
around the magnetic axis requires that $k$ is a positive
integer. Here, we used the polar coordinates $(r,\theta,\varphi)$.
Note that this magnetic field distribution is the Newtonian limit of 
that used in the present GRMHD simulation. In this analysis, we omit
the buoyancy or take $A_i=0$ to focus on the magnetic
instability.

\begin{figure*}[htbp]
\begin{center}
\begin{tabular}{cc}
\includegraphics[scale=0.6]{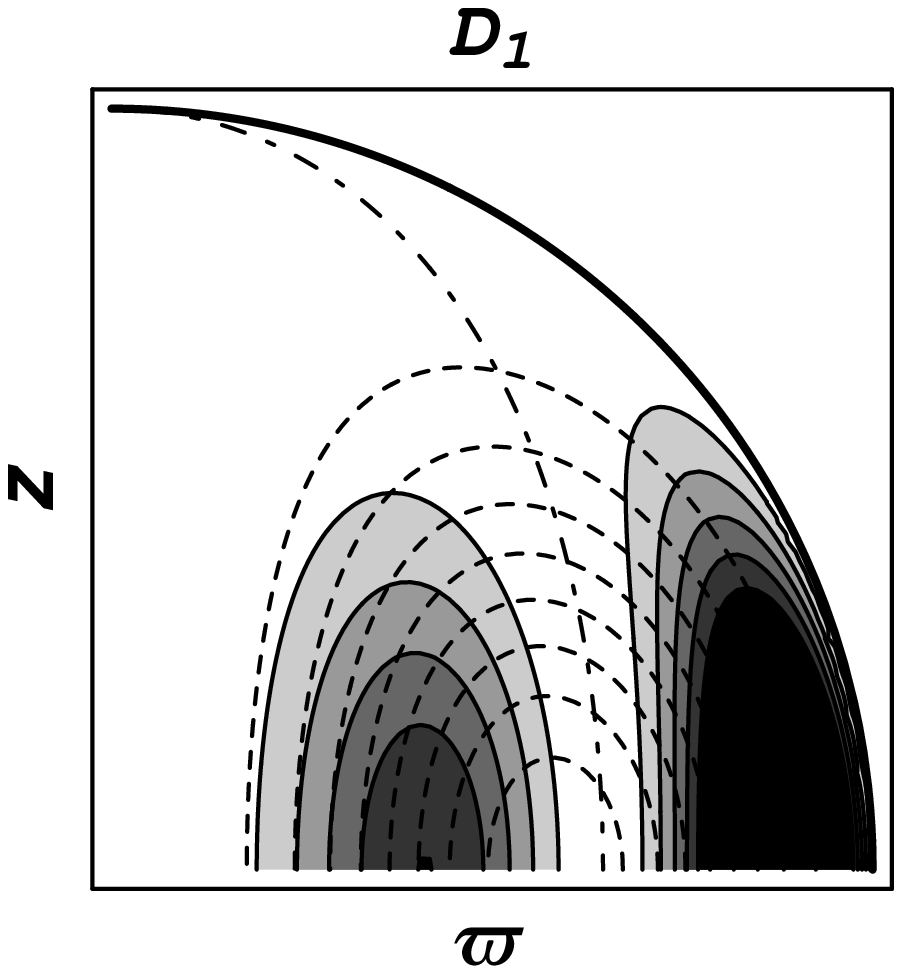} &
~~~~~~~~~~~~\includegraphics[scale=0.6]{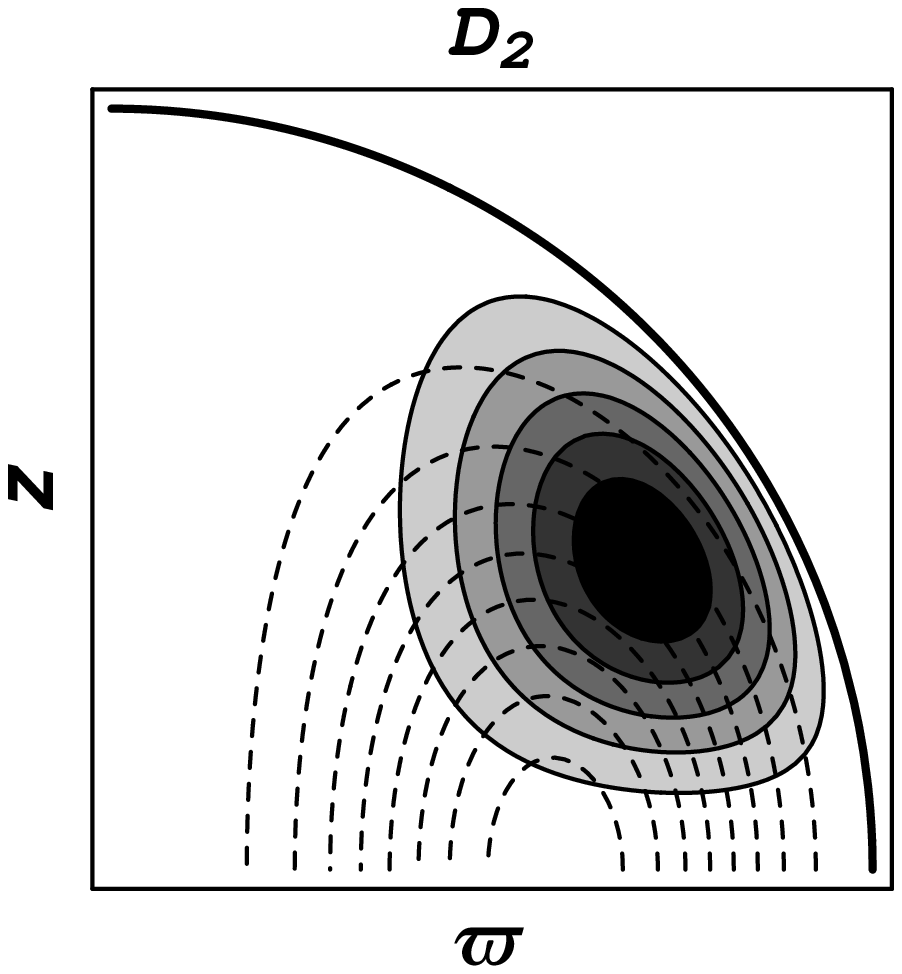} \\
\end{tabular}
\end{center}
\caption{Contours curves (solid curves) of $D_1$ (left) and $D_2$
  (right) on the meridional cross section for the $n=1$ polytropic
  star containing weak toroidal magnetic fields with $k=2$.  The
  darker regions are locally more unstable, whereas the white regions
  correspond to neutrally stable ones (regions with $D_1 \approx 0
  \approx D_2$).  The contours of equi-$D$ are linearly spaced; the
  difference between two adjacent contours is a sixth of the
  difference between the maximum and minimum values of $D_1$ and $D_2$. The
  maximum value is zero and the minimum values are $-4.96$ and $-1.07$
  for $D_1$ and $D_2$, respectively.  The thick quarter circle denotes
  the surface of the star. Inside the star, the dashed curves and the
  dashed-dotted curve show equi-$B$ contours and the critical surface,
  respectively. }
\label{FIG12}
\end{figure*}

First, we pay attention to the nonrotating case. 
Then, the instability conditions (\ref{ineq1}) and
(\ref{ineq2}) are written as 
\beqn
 D_1(r,\theta)&\equiv& {r_0^2\over v_{A,0}^2} 
  \left(g_\varpi-{2v_s^2\over
    r\sin\theta}\right){v_A^2\over v_s^2}
  \left(\sin\theta{\partial\over \partial r}+{\cos\theta\over
    r}{\partial\over \partial\theta}\right) \nonumber \\
&&~~~~~~~  \times \ln\left({B\over\rho\,r\sin\theta}\right)<0\,,\\
   D_2(r,\theta)&\equiv& {r_0^2\over v_{A,0}^2} 
  g_z{v_A^2\over v_s^2}\,
  \left(\cos\theta{\partial\over \partial r}-{\sin\theta\over
    r}{\partial\over \partial\theta}\right) \nonumber \\
&&~~~~~~~ \times  \ln\left({B\over\rho\,r\sin\theta}\right)<0\,, 
\eeqn
Note that by definition, $D_1$ and $D_2$ are dimensionless quantities
and are independent of the magnetic field strength $b_0$.  The
left-hand side of Eq. (\ref{ineq3}) vanishes in the present situation
and this third instability condition cannot give any useful
information.  For the case of $k=1$, it can be seen that $D_1=0=D_2$
because $\partial_i \ln(B/\rho\,r\sin\theta)\propto k-1$. Thus, the
star is neutrally stable for $k=1$. In Fig. \ref{FIG12}, we give the
distributions of $D_1$ and $D_2$ on the meridional cross section for
the case of $k=2$. In this figure, the darker regions have locally
larger growth rates of the unstable mode, whereas the white regions
correspond to neutrally stable ones (regions of $D_1 \approx 0$ and
$D_2 \approx 0$).  We then see that the magnetized stars with $k=2$
are indeed locally unstable. In Fig. \ref{FIG12}, we confirm that the
unstable regions determined by the criterion $D_1$ are separated by
the the critical surface.

The instability determined by $D_1$ occurs primarily near the
equatorial plane and relatively high-density region. By
contrast, the instability determined by $D_2$ occurs near the surface
and for the region of relatively weak magnetic field, and this
indicates that this mode plays a minor role for redistribution of the
magnetic field and for inducing convection. The instability found in
our numerical simulation is likely to be associated with the mode
determined by $D_1$. Hence, in the following, we focus primarily on
this mode. 

The modes associated with $D_1$ and $D_2$ determine the instability in
particular for $l/n \rightarrow 0$ and $l/n \rightarrow \infty$, respectively.
Thus, we focus on the case for small values of $l/n$. As discussed in
\cite{Spruit,Tayler2}, the Tayler instability is associated with a
motion perpendicular to the magnetic axis.  This type of motion
corresponds to the limit of $l/n\rightarrow 0$ in this analysis
because $\delta v^z=-(l/n) \delta v^\varpi$ (see
Eq. (\ref{eq:p_mass_conservation})). Thus, from this point of view, 
it is reasonable to pay attention for the mode associated 
with $D_1$.

%

\begin{figure}[htbp]
\begin{center}
\begin{tabular}{cc}
\includegraphics[scale=0.65]{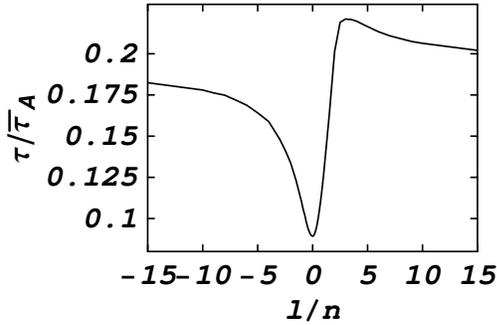} 
\end{tabular}
\end{center}
\vspace{-5mm}
\caption{Growth time $\tau/\bar{\tau}_A$ for the most unstable mode,
  given as a function of $l/n$. }
\label{FIG13}
\end{figure}

Because the Tayler instability occurs for the magnetic field with $k=2$ 
as discussed above, henceforth, we only consider the case of $k=2$. 
For this model, the averaged Alfv{\'e}n speed $\bar{v}_A$ is given by
\begin{eqnarray}
\bar{v}_A&=&\left({\displaystyle \int B^2 d^3x\over \displaystyle 4\pi
  \int \rho\, d^3x} \right)^{1\over 2} \, \nonumber
\\ &=&\left\{{3\over 2800}(315-200\pi^2+32\pi^4) \right\}^{1\over
  2}v_{A,0} \, \nonumber \\ &\approx & 1.25\,v_{A,0} \,.
\label{van}
\end{eqnarray}
where $v_{A,0}$ is defined as
$$
v_{A,0}\equiv {b_0\over\sqrt{4\pi\rho_0}}\,. 
$$
The growth time $\tau$ for the Tayler instability defined in 
Sec. III is associated with increase in the kinetic energy. 
The growth time $\tau$ for the most unstable mode is therefore 
given by 
\begin{eqnarray}
\tau/\bar{\tau}_A=(2\sigma\bar{\tau}_A)^{-1}\approx 0.199 \,
{\rm Min}\left({v_{A,0}\over r_0\,\sigma}\right)\,, 
\end{eqnarray}
where ${\rm Min}(Q)$ denotes the minimum value of $Q(r,\theta)$.  In
the weak magnetic field approximation assumed,
$v_{A,0}(r_0\,\sigma)^{-1}$ is given by
\begin{eqnarray}
{v_{A,0}\over r_0\,\sigma}=\left\{
{1+\left({l\over n}\right)^2\over \left| D_2 \left({l\over n}\right)^2+
\tilde{b}\left({l\over n}\right)+D_1\right|}
\right\}^{1\over 2}\,, 
\end{eqnarray}
where $\tilde{b}=(r_0^2v_{A,0}^{-2})\,b$. In Fig. \ref{FIG13}, we
show the growth time $\tau/\bar{\tau}_A$ obtained by the local
analysis as a function of $l/n$.  As mentioned above, 
we focus on the case that $l/n$ is small. Then, 
Fig. \ref{FIG13} shows that the unstable mode characterized by
$l/n=0$ is the most unstable one, whose growth time is given by
\begin{eqnarray}
\tau/\bar{\tau}_A \approx 0.089\,. 
\end{eqnarray}
Thus, the minimum growth time is by a factor
of 3--5 shorter than that obtained by the GRMHD simulations (compare
with Table II), but the order of magnitude agrees. For modes with a
moderate value of $l/n$, as shown in Fig. \ref{FIG13}, the growth time
$\tau/\bar{\tau}_A$ increases.  For a mode with $l/n\approx 2$, for
example, $\tau/\bar{\tau}_A \approx 0.2$, which is 1/3--2/3 of the
growth time shown in Table II.  This result is reasonable because we
here assume a Newtonian model whereas in the simulations, we adopt
highly general relativistic neutron stars for which 
the profiles of density and magnetic field are significantly different
from the Newtonian ones.

\begin{figure}[htbp]
\begin{center}
\begin{tabular}{cc}
\includegraphics[scale=0.65]{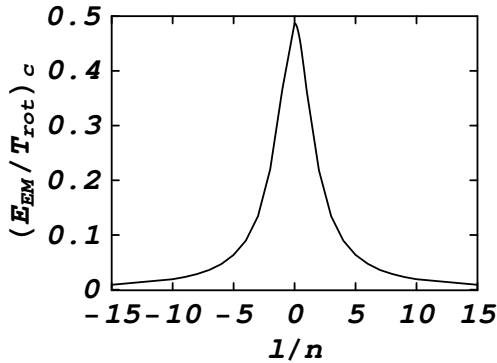}
\end{tabular}
\end{center}
\vspace{-5mm}
\caption{Critical ratio of the electromagnetic energy to the
  rotational kinetic energy, $\left(E_{\rm EM}/T_{\rm rot}\right)_c$,
  given as a function of $l/n$. }
\label{FIG14}
\end{figure}
Next, we consider the slowly rotating model. In the slow
rotation and weak magnetic field approximation, the 
criterion of the Tayler instability becomes
\begin{eqnarray}
&&{r_0^2\Omega^2\over v_{A,0}^2} + {1\over 4} \left\{D_2
  \left({l\over n}\right)^2+ \tilde{b}\left({l\over
    n}\right)+D_1\right\} < 0 \,.
\label{eq:Tayler_inst_condition}
\end{eqnarray}
For the $k=2$ model, the rotational kinetic energy $T_{\rm rot}$ and 
the electromagnetic energy $E_{\rm EM}$ are, respectively,  given by
\begin{eqnarray}
T_{\rm rot}&=&{1\over 2}\int \rho\,r^2\sin^2\theta\Omega^2\, d^3x 
\,\nonumber \\
        &=& {4\over 3}\pi^2(\pi^2-6)\rho_0r_0^5\Omega^2 \,\nonumber \\
        &\approx& 50.9\,\rho_0r_0^5\Omega^2 \,, \\
E_{\rm EM}&=&{1\over 8\pi}\int B^2\, d^3x \, \nonumber \\
         &=& {3\pi\over 5600}(315-200\pi^2+32\pi^4)r_0^3 b_0^2\, \nonumber \\
        &\approx& 2.45\,r_0^3 b_0^2 \,.  
\end{eqnarray}
The ratio of the electromagnetic energy to the rotational kinetic 
energy is then written as 
\begin{eqnarray}
E_{\rm EM}/T_{\rm rot}\approx 0.605\,{v_{A,0}^2\over r_0^2\Omega^2}\,.  
\end{eqnarray}
In terms of $E_{\rm EM}/T_{\rm rot}$, thus, 
Eq. (\ref{eq:Tayler_inst_condition}) is rewritten as  
\begin{eqnarray}
E_{\rm EM}/T_{\rm rot}& > & \left(E_{\rm EM}/T_{\rm rot}\right)_c \,\nonumber \\
&\equiv& {\rm Min}\left({-2.42\over D_2 \left({l\over n}\right)^2+
\tilde{b}\left({l\over n}\right)+D_1 }\right) \,. 
\label{rat_T_E}
\end{eqnarray}
For the polytropic models with $n=1$ and $k=2$, numerical values of 
$ \left(E_{\rm EM}/T_{\rm rot}\right)_c$ are shown as a function of $l/n$ 
in Fig. \ref{FIG14}.

Form this figure, it is found that for the most unstable mode, whose value of $l/n$ is zero, 
the Tayler instability sets in if the condition,
\begin{eqnarray}
E_{\rm EM}/T_{\rm rot} > 0.49 \,, 
\label{local_stability_condition}
\end{eqnarray}
is satisfied.  For modes with a moderate value of $l/n$, values of
$\left(E_{\rm EM}/T_{\rm rot}\right)_c$ decreases, e.g.  $\left(E_{\rm
  EM}/T_{\rm rot}\right)_c\approx 0.21$ for a mode with $l/n\approx
2$.  As argued in Sec. III, the instability condition obtained by the
GRMHD simulation is $E_{\rm EM}/T_{\rm rot} \agt 0.2$, which is the
same order as that of Eq. (\ref{local_stability_condition}).

For the small values of $E_{\rm EM}/T_{\rm rot}$, the unstable modes
should have larger values of $l/n$, and correspondingly, the growth
time scale becomes longer. This suggests that even if a model star
appears to be stable in a numerical simulation for a finite duration,
the star might become unstable for a sufficiently long run. 

As found from Fig. \ref{FIG14}, magnetized star is always unstable for
modes with $|l/n|\approx\infty$ irrespective of the rotation rate. As
mentioned previously, however, these modes are associated with
$D_2$. This instability occurs near the stellar surface and its effect
would not be significant for global redistribution of the magnetic
field profile. Thus, although all the magnetized stars with $k=2$ are
unstable strictly speaking, this type of instability does not seem to
play a significant role.

\end{document}